\documentclass[manuscript]{aastex}
\usepackage{emulateapj5}
\usepackage{apjfonts}

\font\mitt=cmmi9 scaled\magstep 1

\slugcomment{The Astrophysical Journal, accepted Oct 2, 2002}

\def\gtsima{$\; \buildrel > \over \sim \;$}
\def\ltsima{$\; \buildrel < \over \sim \;$}
\def\gsim{\lower.7ex\hbox{\gtsima}}
\def\lsim{\lower.7ex\hbox{\ltsima}}
\def\simgt{\lower.7ex\hbox{\gtsima}}
\def\simlt{\lower.7ex\hbox{\ltsima}}
\def\la{\lsim}
\def\ga{\gsim}
\def\lta{\la}
\def\gta{\ga}

\def\colhead#1{#1}

\def\HI{\ifmmode \hbox{\scriptsize H\kern0.5pt{\footnotesize\sc i}}\else H\kern1pt{\small I}\fi}
\def\Halpha{H$\alpha$}
\def\mlstar{\ifmmode\Upsilon_{\!\!*}\else$\Upsilon_{\!\!*}$\fi}
\def\mlstarR{\ifmmode\Upsilon_{\!\!*}^R\else$\Upsilon_{\!\!*}^R$\fi}
\def\kms{\ifmmode\mathrm{~km~}\mathrm{s}^{-1}\else km s$^{-1}$\fi}
\def\cm2{cm$^{-2}$}
\def\pc2{pc$^{-2}$}
\def\pc3{pc$^{-3}$}
\def\varv{\hbox{\mitt v}}
\def\chisq{$\chi^2$}
\def\rchisq{$\chi^2_r$}
\def\Msun{M$_\odot$}
\def\Lsun{L$_\odot$}
\def\lab{\rlap{\raise2pt\hbox{$<$}}\lower2.5pt\hbox{$\sim$}}
\def\gab{\rlap{\raise2pt\hbox{$>$}}\lower2.5pt\hbox{$\sim$}}

\newcommand{\tskip}{\omit\tablevspace{1pt}}

\shorttitle{The Central Mass Distribution in Dwarf and LSB Galaxies}
\shortauthors{Swaters, Madore, van den Bosch \& Balcells}

\begin{document}

\title{The Central Mass Distribution in Dwarf and Low Surface
  Brightness Galaxies}

\author{R. A. Swaters}

\affil{Department of Physics and Astronomy, Johns Hopkins University,
  3400 N. Charles Str., Baltimore, MD 21218, Space Telescope
  Science Institute, 3700 San Martin Dr., Baltimore, MD 21218, and
  Department of Terrestrial Magnetism, Carnegie Institution of
  Washington, 5241 Broad Branch Rd NW, Washington, DC 20015 }
\medskip

\author{B. F. Madore}
\affil{Observatories of the Carnegie Institution of Washington and
  NASA/IPAC Extragalactic Database, California Institute of
  Technology, Pasadena CA 91125, USA}

\medskip

\author{Frank C. van den Bosch}
\affil{Max Planck Institut f\"ur Astrophysik, Karl Schwarzschild
Str. 1, Postfach 1317, 85741 Garching, Germany}

\and

\vskip-6pt
\author{M. Balcells}
\affil{Instituto de Astrof\'{\i}sica de Canarias, E-38200 La Laguna,
Tenerife, Spain}

\begin{abstract}
  We present high-resolution H$\alpha$ rotation curves for a sample of
  15 dwarf and low surface brightness galaxies.  From these, we derive
  limits on the slopes of the central mass distributions, using both a
  direct inversion of the rotation curves as well as detailed mass
  models.  Assuming the density distributions of dark matter halos
  follow a power-law at small radii, $\rho(r)\propto r^{-\alpha}$, we
  find inner slopes in the range $0 \la \alpha \la 1$ for most
  galaxies.  Thus, even with the relatively high spatial resolution of
  the H$\alpha$ rotation curves presented here the inner slopes are
  poorly constrained.  In general, halos with constant density cores
  ($\alpha=0$) provide somewhat better fits, but the majority of our
  galaxies ($\sim 75$ percent) are also consistent with $\alpha = 1$,
  as long as the $R$-band stellar mass-to-light ratios are smaller
  than about 2.  Halos with $\alpha = 1.5$, however, are ruled out in
  virtually every case.  In order to investigate the robustness of
  these results we discuss and model several possible causes of
  systematic errors including non-circular motions, galaxy
  inclination, slit width, seeing, and slit alignment errors. Taking
  the associated uncertainties into account, we conclude that even for
  the $\sim 25$ percent of the cases where $\alpha=1$ seems
  inconsistent with the rotation curves, we cannot rule out cusp
  slopes this steep. Inclusion of literature samples similar to the
  one presented here leads to the same conclusion when the possibility
  of systematic errors is taken into account.  In the ongoing debate
  on whether the rotation curves of dwarf and low surface brightness
  galaxies are consistent with predictions for a cold dark matter
  universe, we argue that our sample and the literature samples
  discussed in this paper provide insufficient evidence to rule out
  halos with $\alpha=1$. At the same time, we note that none of the
  galaxies in these samples require halos with steep cusps, as most
  are equally well or better explained by halos with constant density
  cores.
\end{abstract}

\keywords{ galaxies: dwarfs --- galaxies: halos --- galaxies:
  kinematics and dynamics }

\section{Introduction} 
\label{secintro}

Cosmological simulations based on different properties of dark matter
make distinct predictions for the properties of the dark matter halos.
The inner slope $\alpha$ of the central power-law density distribution
$\rho(r)\propto r^{-\alpha}$ is particularly sensitive to the adopted
properties of dark matter.  If, for example, the dark matter is
assumed to be cold and collisionless (CDM), the equilibrium density
profiles of the dark halos will have steep inner slopes. Early
simulations found $\alpha=1$ (Dubinski \& Carlberg 1991, Navarro,
Frenk, \& White 1996, 1997), but later simulations found steeper
slopes with $\alpha=1.5$ (Fukushige \& Makino 1997; Moore et al. 1998,
1999; Dav\'e et al. 2001; Klypin et al. 2001). A recent comprehensive
study of the effects of numerical parameters by Power et al. (2002)
places an upper limit on the inner slope of $\alpha=1.2$, and suggests
that the disagreement in the literature on the value of $\alpha$ is
mainly due to resolution issues in the numerical simulations.  If the
dark matter is assumed to be warm (WDM), cores with $\alpha\sim1$ are
found (Col{\'\i}n, Avila-Reese, \& Valenzuela 2000; Knebe et
al. 2002), and shallower slopes are found for self-interacting dark
matter (SIDM, Dav\'e et al.  2001).  Because of these differences in
inner slopes it may be possible to obtain limits on the nature of dark
matter from the observed rotation curves of disk galaxies.

Unfortunately, it has proved remarkably hard to establish the inner
slope of the dark matter distribution observationally.  The inner
slope is usually constrained by fitting the observed rotation curves
with mass models that contain both the dark and luminous components.
However, because the mass-to-light ratio (M/L) of the stellar disk is
not known, a large degeneracy exists in the mass modeling (e.g., van
Albada et al. 1985; Lake \& Feinswog 1989). As a result, the observed
rotation curves can often be explained by a wide range in mass models,
ranging from mass models in which the stellar disk dominates in the
inner parts and where the halo has a shallow inner profile (e.g.,
Begeman 1987, 1989), to ones in which the stellar disk is less
important and the mass distribution is dominated by a centrally
concentrated halo (e.g., Verheijen 1997; Navarro 1998).

\begin{table*}[t]
\begin{center}
{\sc Table \ref{tabglobpars}\\ \smallskip\hbox to\hsize{\hfil{Global properties}\hfil}}
\small
\setlength{\tabcolsep}{6pt}
\begin{tabular}{lrrrrrrrrcrrrrc}
\tskip \tableline
\tableline \tskip
\colhead{Name} & \multicolumn{3}{c}{R.A. (J2000)} & \multicolumn{3}{c}{Dec. (J2000)} &
\colhead{$M_R$} & \colhead{$h$} & \colhead{$\mu_R$} & \colhead{$D$} &
\colhead{$\varv_{\rm sys}$} & \colhead{$i$} & \colhead{$\phi$} & \colhead{notes}\\
\colhead{} & \multicolumn{3}{c}{} & \multicolumn{3}{c}{} &
\colhead{(mag)} & \colhead{(kpc)} & \colhead{(mag $\prime\prime^{-2}$)} & \colhead{(Mpc)} & \colhead{(\kms)} & \colhead{($^\circ$)} & \colhead{($^\circ$)} & \\
\colhead{(1)} & \multicolumn{3}{c}{(2)} & \multicolumn{3}{c}{(3)} & \colhead{(4)} & \colhead{(5)} & \colhead{(6)} & \colhead{(7)} & \colhead{(8)} & \colhead{(9)} & \colhead{(10)} & \colhead{(11)} \\
\tableline \tskip
UGC 731 & 1 & 10 & 43.6 & 49 & 36 & 4 & -16.6 & 1.65 & 23.0 & 8.0 & 638 & 57 & 77 & barred \\
UGC 2259 & 2 & 47 & 55.6 & 37 & 32 & 17 & -17.8\rlap{$^a$} & 1.29\rlap{$^a$} & 21.2\rlap{$^a$} & 9.8 & 574 & 41 & 160 & \\
UGC 4325 & 8 & 19 & 19.7 & 50 & 0 & 32 & -18.1 & 1.63 & 21.6 & 10.1 & 527 & 41 & 53 & \\
UGC 4499 & 8 & 37 & 41.4 & 51 & 39 & 9 & -17.8 & 1.49 & 21.5 & 13.0 & 693 & 50 & 140 & barred \\
UGC 5721 & 10 & 32 & 17.3 & 27 & 40 & 8 & -16.6 & 0.45 & 20.2 & 6.7 & 535 & 61 & 96 & \\
UGC 8490 & 13 & 29 & 36.4 & 58 & 25 & 12 & -17.3 & 0.66 & 20.5 & 4.9 & 208 & 50 & 170 & \\
UGC 11557 & 20 & 24 & 0.7 & 60 & 11 & 41 & -19.7 & 3.10 & 21.0 & 23.8 & 1388 & 30 & 94 & \\
UGC 11707 & 21 & 14 & 31.7 & 26 & 44 & 5 & -18.6 & 4.30 & 23.1 & 15.9 & 904 & 68 & 57 & \\
UGC 11861 & 21 & 56 & 24.2 & 73 & 15 & 39 & -20.8 & 6.06 & 21.4 & 25.1 & 1481 & 50 & 39 & barred \\
UGC 12732 & 23 & 40 & 39.8 & 26 & 14 & 11 & -18.0 & 2.21 & 22.4 & 13.2 & 749 & 39 & 14 & \\
F 563-V2 & 8 & 53 & 3.7 & 18 & 26 & 9 & -19.0\rlap{$^b$} & 2.10\rlap{$^b$} & 21.3\rlap{$^b$} & 61 & 4310 & 29 & 148 & barred \\
F 568-1 & 10 & 26 & 6.3 & 22 & 26 & 1 & -18.9 & 4.41 & 22.7 & 85 & 6524 & 26 & 13 & \\
F 568-3 & 10 & 27 & 20.3 & 22 & 14 & 22 & -19.2 & 4.48 & 22.4 & 77 & 5911 & 40 & 169 & barred \\
F 568-V1 & 10 & 45 & 2.1 & 22 & 3 & 16 & -18.7 & 3.96 & 22.7 & 80 & 5769 & 40 & 136 & \\
F 574-1 & 12 & 38 & 7.1 & 22 & 18 & 50 & -19.2 & 4.24 & 22.2 & 96 & 6889 & 65 & 90 & \\
\tableline
\end{tabular}
\end{center}

\centerline{\vbox{\hsize=15.5cm \footnotesize 
Notes --- (1) the name of the galaxy, (2) and (3) the
coordinates, (4) the absolute $R$-band magnitude, (5) the scale
length, (6) the $R$-band extrapolated central disk surface brightness,
(7) the adopted distance, based on a Hubble constant of 75 \kms
Mpc$^{-1}$ or on a secondary distance indicator (see Swaters \&
Balcells 2002) (8) the \HI\ heliocentric systemic velocity, (9) the
inclination, (10) the position angle of the major axis, and (11) notes
on galaxy morphology, as judged from deep CCD imaging presented in
Swaters \& Balcells (2002) and de Blok et al. (1995). The data for
UGC~2259 have been taken from Carignan, Sancisi, \& van Albada 1988,
the data for the remaining UGC galaxies comes from Swaters 1999, the
data for the LSB galaxies comes from McGaugh \& Bothun 1994 and de
Blok et al. 1995, 1996.\break $^a$ Converted from $I$-band observations
assuming $R-I=0.5$. $^b$ Converted from $B$-band observations assuming
$B-R=0.8$.}}

\end{table*}

This degeneracy can be partially avoided by studying dwarf and low
surface brightness (LSB) galaxies. Even though the central parts of
the observed rotation curves can in principle be explained by scaling
up the contribution of the stellar disks (Swaters 1999, hereafter S99;
Swaters, Madore, \& Trewhella 2000, hereafter SMT), the inferred
stellar M/Ls are much higher than expected from population synthesis
modeling (e.g., Worthey 1994; Bell \& de Jong 2001).  For reasonable
stellar M/Ls, these galaxies are then dominated by dark matter at all
radii, making dwarf and LSB galaxies very suitable for studies of the
properties of dark matter.

Consequently, the rotation curves of these galaxies have received a
great deal of attention.  Early studies found that these galaxies have
slowly rising inner rotation curves and that their halos can be well
described by pseudo-isothermal spheres with constant density cores
(e.g., Carignan \& Beaulieu 1989; C\^ot\'e, Carignan, \& Sancisi 1991;
Broeils 1992a, 1992b; C\^ot\'e 1995; de Blok \& McGaugh 1997).
Because of their shallow slopes, these rotation curves were stated to
be inconsistent with the steep halos predicted by cold dark matter
(Moore 1994; Flores \& Primack 1994; Navarro et al. 1996; McGaugh \&
de Blok 1998). Recent studies raised the concern that some of these
\HI\ rotation curves may have been affected by beam smearing due to
the poor angular resolution of the \HI\ observations, and found that
the rotation curves of dwarf and LSB galaxies rise more steeply when
beam smearing is taken into account (S99; Blais-Ouellette et al. 1999;
SMT).  Fitting beam-smeared mass models to the \HI\ rotation curves
indicated that dwarf and LSB galaxies are consistent with a wide range
of dark matter properties, ranging from constant density cores to
steep cusps with $\alpha=1$. However, slopes with $\alpha=1.5$ appear
difficult to reconcile with the observations (van den Bosch et al.
2000; Swaters 2001a; van den Bosch \& Swaters 2001, hereafter vdBS).

\begin{figure*}[th]
\begin{center}
\epsfxsize=1.0\hsize\epsfbox{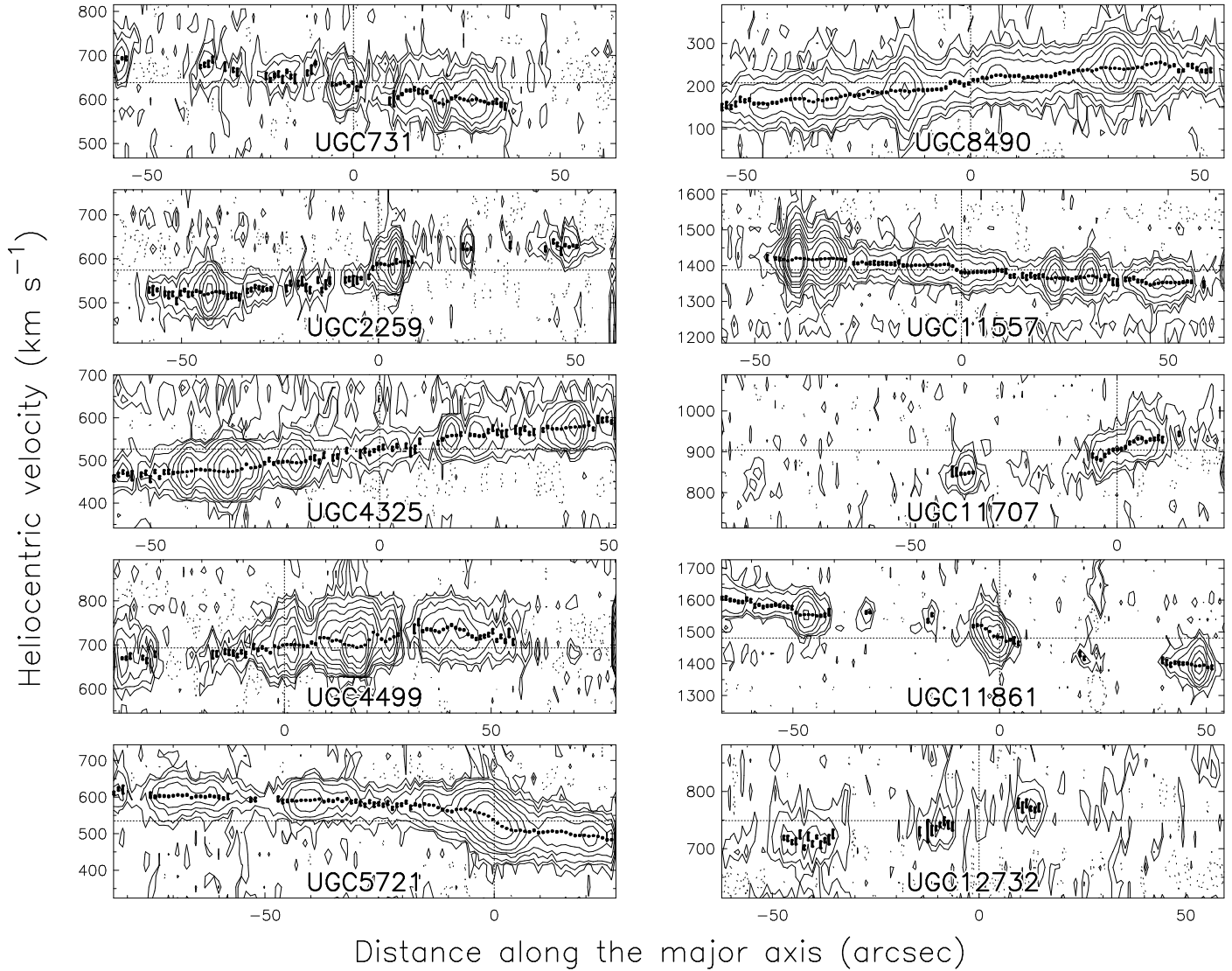}
\caption{\Halpha\ position-velocity diagrams for the 10 dwarf galaxies
in our sample. The spectra have been binned to $1''$ to increase the
signal-to-noise ratio. Contour levels are -2, 2, 4, 8, 16, ...,
negative contours are dotted. The dots with error bars give the radial
velocities with their formal errors as derived from Gauss fits to the
velocity profiles. The vertical dotted lines indicate the galaxy
centers, the horizontal dotted lines indicate the systemic velocities.
\label{figdata}}
\end{center}
\end{figure*}

\HI\ observations, however, with typical resolutions of $15''$, may
not be best suited to determine the inner rotation curve slopes. The
large range in mass models that are consistent with the observations
might simply be a reflection of the systematic effects caused by the
\HI\ observations. High angular resolution observations, for example
based on \Halpha\ long slit or Fabry-Perot spectroscopy seem more
suited to measure the inner slopes. Yet even among studies based on
such high angular resolution observations a controversy remains.  On
the one hand, there are studies that find that dwarf and LSB galaxies
may have steep inner slopes (Swaters 2001b; Pickering et al., in
preparation), and on the other hand there are several studies that find that
these galaxies are not consistent with steep inner slopes (Borriello
\& Salucci 2001; Dalcanton \& Bernstein 2000; Blais-Ouellette, Amram,
\& Carignan 2001; de Blok et al. 2001a, hereafter dBMBR; de Blok
et al. 2001b, hereafter dBMR; de Blok \& Bosma 2002, hereafter dBB;
Marchesini et al. 2002).

This paper presents new \Halpha\ long slit spectroscopy for a sample
of 15 dwarf and LSB galaxies, and directly addresses the halo cusp
slope controversy outlined above.  We measure the inner slopes of the
mass distributions, and determine the range of mass models consistent
with the observations, taking account of possible systematic
effects. In Section~\ref{secsample} we describe the sample and the
observations.  Section~\ref{secderrcs} covers the derivation of the
rotation curves.  In Section~\ref{secmassmodels} we describe the mass
models fitted to the rotation curves, and in Section~\ref{secresults}
we present the results from the mass modeling.  In
Section~\ref{secobseffs} the systematic effects that may have affected
the derived rotation curves and inner slopes are discussed, and in
Section~\ref{secmodeling} we model how these systematic effects affect
the data. In Section~\ref{secdisc} we discuss our results, and we
compare our results to those in the literature, and to predictions
made by cosmological simulations.  In Section~\ref{secconcl} we
present our conclusions.

\section{Sample and observations}
\label{secsample}

The dwarf galaxies presented here were taken from the sample of S99,
and the LSB galaxies were taken from the sample of de Blok, McGaugh,
\& van der Hulst (1996). The UGC galaxies in our sample are dwarf
galaxies, the remaining five galaxies are LSB galaxies. We note that
the properties of dwarf and LSB galaxies are fairly similar. Like LSB
galaxies, dwarf galaxies usually have low surface brightnesses, but
they may have high surface brightnesses as well (e.g., UGC~5721 and
UGC~8490 in the sample presented here), and dwarf galaxies usually
have smaller scale lengths than LSBs, but the selection criteria used
by S99 resulted in the inclusion of galaxies with properties similar
to the LSB galaxies in de Blok et al.  (1996).

The observations were carried out with the Double Spectrograph at the
Palomar Observatory with the $200''$ Hale telescope, spread out over
different observing runs, on March 2 and 3, 1997, October 26 to 28,
1998 and November 20, 1998.  The March 1997 run was for a different
project, and only 3 galaxies were observed from the S99 sample.
Because of adverse weather conditions during the other runs, only 12
more galaxies could be observed from the S99 and de Blok (1996)
samples. The observed galaxies and their global properties are listed
in Table~\ref{tabglobpars}.

The slit width for all observations was $1''$, the FWHM velocity
resolution and the pixel size in the spatial direction are $54 \kms$
and $0.5''$, respectively.  All galaxies were observed in a
single 1800s exposure with the slit oriented along the major axis, at
the position angle listed in Table~\ref{tabglobpars}.  The LSB
galaxies all appeared to have a well-defined nucleus, making it
possible to align the slit with the center of the galaxy by eye.  For
the dwarf galaxies, the alignment of the slit proved more difficult,
because of the diffuse, extended nature of these objects.  The
position of the slit was judged in relation to the galaxy image and to
stars in the field of view.  The systematic effects that may have been
introduced by an incorrectly positioned slit are discussed in
Section~\ref{secobseffs}.

\begin{figure*}
\epsfxsize=1.0\hsize\epsfbox{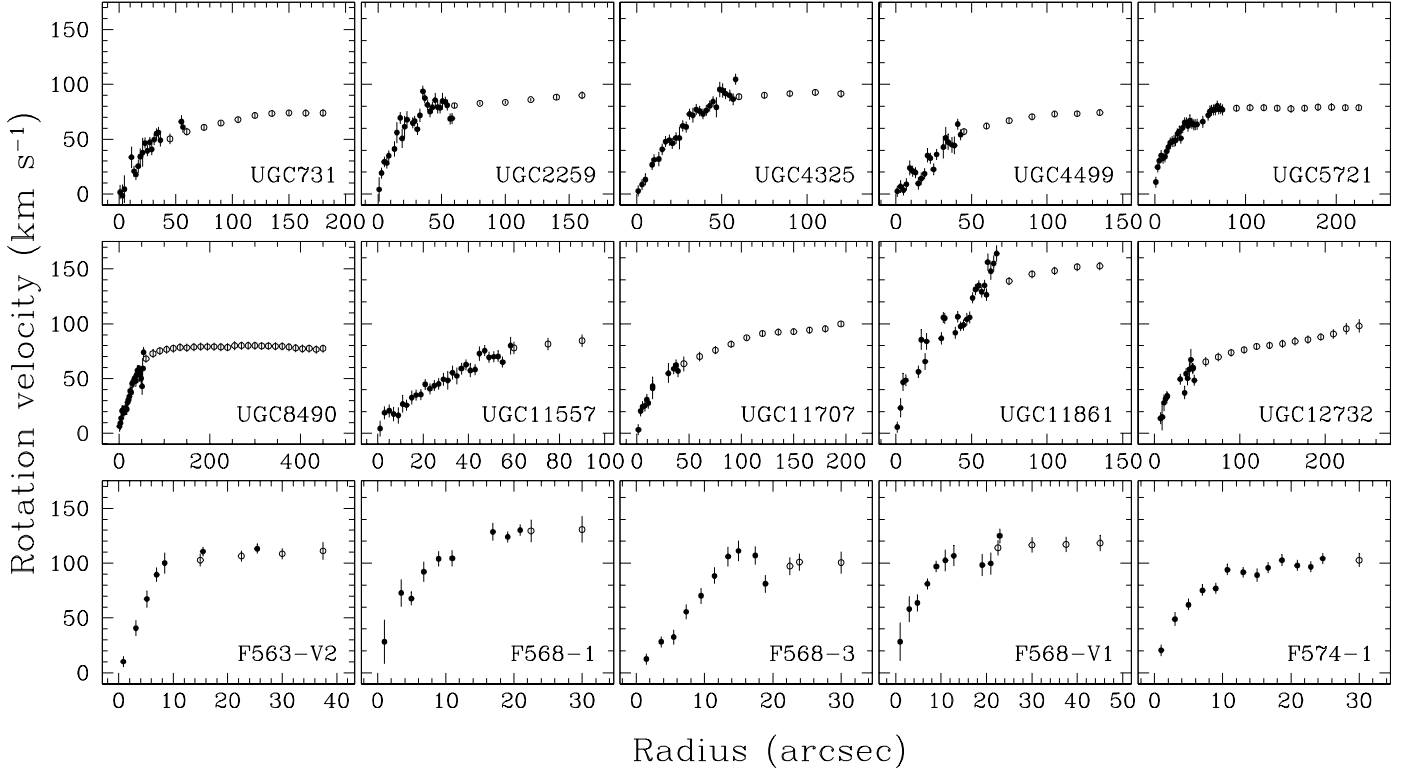}
\caption{Combined \Halpha/\HI\ rotation curves. The filled circles
represent the \Halpha\ rotation curves as derived in this paper, the
open circles are the \HI\ rotation curves from Swaters (1999) for the
dwarf galaxies and from de Blok et al. (1996) for the LSB galaxies.
\label{figrcs}}
\end{figure*}

\section{Derivation of the rotation curves}
\label{secderrcs}

The data were bias subtracted, flatfielded, cleaned from cosmic ray
events, wavelength calibrated, sky subtracted and continuum subtracted
using standard procedures in {\sc iraf}.  To increase the
signal-to-noise ratio, we rebinned the spectra to $1''$ pixels.  The
resulting \Halpha\ position-velocity diagrams are shown in
Fig.~\ref{figdata}.  The LSB galaxies have already been presented in
SMT and are not repeated here.  To determine the radial velocities, we
fit Gaussians to the line profiles at each position where the peak
emission was higher than 3 times the noise level.  These fits and
their formal errors are overlayed on the \Halpha\ position-velocity
diagrams in Fig.~\ref{figdata}.

The position of the galaxy center along the slit was determined from
the peak in the continuum.  For most galaxies the continua were
sufficiently bright and compact that the center could be determined to
an accuracy better than $1''$.  For a few galaxies for which the
continuum was less peaked and/or fainter the position of the center
was first estimated from the broad continuum peak and then refined by
minimizing the differences between the receding and approaching sides
of the rotation curves.  The resulting shifts in the position of the
center were at most a few arcseconds.  The systemic velocities were
determined at the position of the center and have been corrected to
heliocentric velocities.  In all cases they agree with the \HI\
systemic velocities within the errors.

To determine the rotation curves, the radial velocities as determined
from the Gaussian fits were folded around the center and systemic
velocity.  The velocity points were binned in intervals of $2''$
(similar to the angular resolution of our observations) and averaged
both in radius and velocity, weighted with the inverse square of the
errors.  The final error on each point of the rotation curve was
defined ad hoc as half the quadratic sum of the average error of the
points in the bin, weighted by the inverse square of the errors, and
half the difference between the minimum and maximum rotation velocity
within the bin. A minimum error of $5\kms$ was imposed.

The method to derive the rotation curves as described above is
somewhat different from the method used in SMT, where the rotation
velocity at each interval was estimated by eye, resulting in
relatively smooth rotation curves.  We have reanalyzed the SMT data
with the procedure described above, so that all rotation curves
presented in this paper are determined in the same way. The method
used here avoids any bias or artificial smoothness that may have been
introduced by determining the rotation curve by eye.

To extend the radial coverage of the rotation curves derived from the
\Halpha\ observations, we combined them with the \HI\ rotation curves
from Carignan et al. (1988) for UGC~2259, S99 for other dwarf
galaxies, and from de Blok et al.\ (1996) for the LSB galaxies.  For
the LSB galaxies there are systematic differences between the \HI\ and
\Halpha\ rotation curves, in the sense that the \HI\ rotation curves
appear to underestimate the rotation velocities in the inner parts
(SMT, but see McGaugh, Rubin, \& de Blok 2001), but the rotation
velocities in the outer parts generally agree well.  Hence, we
replaced the inner parts of the \HI\ rotation curves with the \Halpha\
data derived here.  The resulting rotation curves are shown in
Fig.~\ref{figrcs}

In principle, the observed rotation curves need to be corrected for
the effects of pressure, which may provide part of the support against
gravity.  However, as discussed in more detail in S99, these corrections are
uncertain and usually small, especially in the galaxy centers.
Therefore, we did not correct the observed rotation curves for
asymmetric drift and we assumed that the observed rotation curves are
a good representation of the circular velocities.

\section{Mass modeling}
\label{secmassmodels}

Assuming that  a galaxy is  axially symmetric and in  equilibrium, the
circular velocity directly reflects the total gravitational potential
\begin{equation}
-F_r = \frac{d\Phi}{dr} = \frac{{\varv_c}^2}{r},
\label{eqpot}
\end{equation}
where $F_r$  is the radial force, $\Phi$  the gravitational potential,
$r$  the galactocentric  radius and  $\varv_c$ the  circular velocity.
The  total gravitational  potential is  the sum  of  the gravitational
potentials of  the individual mass  components in a galaxy.   Here, we
assume that  the galaxy consists  of three main components:  a stellar
disk, a  gaseous disk, and a  spherical dark halo.  Its total circular
velocity is then given by:
\begin{equation}
{\varv_c} = \sqrt{\mlstar{\varv_d}^2 +
\eta{\varv_\mathrm{\HI}}^2 + {\varv_h}(p_1,\ldots,p_n)^2},
\label{eqfit}
\end{equation}
where \mlstar\ is the stellar mass-to-light ratio, $\varv_d$ is the
rotation curve of the stellar disk for a stellar mass-to-light ratio
of unity, $\eta$ represents the inclusion of the contribution of
helium to the gaseous component, assumed to be 1.32,
$\varv_\mathrm{\HI}$ is the rotation curve of the \HI\ only, and
$\varv_h(p_1,\ldots,p_n)$ represents the dark halo, where $p_1$ to
$p_n$ are parameters describing its mass distribution.  Each of the
components in this equation is described in more detail below.  The
best fitting mass model for a given dark halo model is determined by
fitting Eq.~\ref{eqfit} to the observed rotation curve, with \mlstar\
and $p_1$ to $p_n$ as free parameters.

\subsection{The stellar disk}

The circular velocities due to the stellar disks have been calculated
from the observed $R$-band luminosity profiles using the method of
Casertano (1983) and Begeman (1987).  For F563-V2 a $B$-band light
profile was used.  The light profiles for the dwarf galaxies have been
taken from Swaters \& Balcells (2002), and for the LSB galaxies from
de Blok, van der Hulst, \& Bothun (1995) and McGaugh \& Bothun
(1994). Throughout we assume that the stellar disk have a constant
mass-to-light ratio \mlstarR\, and follow a sech$^2$ vertical density
distribution with $z_0=h/6$.  This particular choice for the vertical
density distribution has a negligible effect on the resulting rotation
curves.  Finally, the surface brightness profiles of most of the
galaxies in our sample are well described by a pure exponential with
little or no central concentration of light, and we therefore did not
include a bulge component.

\subsection{The gaseous disk}

Circular velocity contributions from the gas disks have been computed
from the radial \HI\ surface density profiles presented in Swaters et
al. (2002) and de Blok et al. (1996). All dwarf galaxies were
sufficiently resolved to allow a direct determination of the radial
\HI\ surface density profile by azimuthally averaging in concentric
ellipses.  The LSB galaxies, however, were poorly resolved, and the
radial profiles derived in this way may be affected by beam smearing
(van den Bosch et al. 2000).  To calculate the radial distribution of
the \HI\ for the LSB galaxies, we have used the algorithm described by
Warmels (1988) that uses a iterative deconvolution scheme (Lucy 1974)
to correct for the effects of beam smearing.  Although the resulting
density profiles may be very different, the effects on mass modeling
are minor because the contribution of the \HI\ to the rotation
velocity at each radius is usually a modest fraction of the total
rotation velocity.  We assumed that the \HI\ layer is infinitely thin.
The choice of the vertical distribution has little effect on the
resulting rotation curve. Even if the gaseous disk were to have the
same thickness as the stellar disk, the difference would still be less
than 5\%.

\subsection{The dark halo}

We consider two different density profiles for the dark halo: a
pseudo-isothermal sphere, and a more general density distribution with
variable central cusp slope.

The pseudo-isothermal sphere has frequently been used in the
literature to describe the dark halo and is able to explain the
observed rotation curves for a wide range of Hubble types and a wide
range of stellar M/Ls (e.g., Begeman 1987, 1989; Broeils 1992a; de
Blok \& MgGaugh 1997; Verheijen 1997; S99). Its density distribution
is given by
\begin{equation}
\rho(r)=\rho_0\left[1+\left(\!\frac{r}{\,r_c}\right)^{\!\!2}\:\right]^{-1},
\label{eqisorad}
\end{equation}
where $\rho_0$ is the central dark matter density, and $r_c$ the core
radius. This density profiles gives rise to a rotation curve of the form
\begin{equation}
\varv_h(r)=\sqrt{4\pi\mathrm{G}\rho_0r_c^{\,2}\left[1-\frac{r_c}{\!r}
\mathrm{arctan}\left(\!\frac{r}{\,r_c}\right)\right]}.
\label{eqisovrot}
\end{equation}

As an alternative density distribution we consider
\begin{equation}
\label{eqgennfw}
\rho(r) = {\rho_0 \over (r/r_s)^{\alpha} (1 + r/r_s)^{3-\alpha}}.
\end{equation}
This density distribution changes from $\rho \propto r^{-\alpha}$ for
$r \ll r_s$ to $\rho \propto r^{-3}$ for $r \gg r_s$. For $\alpha=0$
the density profile thus has a constant density core and becomes
comparable to that of the pseudo-isothermal sphere (at least at small
radii), while for $\alpha=1$ it reduces to the NFW profile (Navarro
et al. 1997).  This `generalized' NFW density distribution has
previously been used in rotation curve analyses by e.g., van den Bosch
et al. (2000) and vdBS. The corresponding circular velocity curve is
\begin{equation}
\label{eqrcgen}
\varv_h(r)=\varv_{200}\sqrt{{\mu(xc)}\over{x\mu(c)}},
\end{equation}
where $c = r_{200}/r_s$, with $r_{200}$ the radius inside of which the
mean  density is  200  times  the critical  density  for closure,  
$x=r/r_{200}$, and
\begin{equation}
\label{eqmu}
\mu(x)=\int\limits_0^xy^{2-\alpha}(1+y)^{\alpha-3}{\rm d}y.
\end{equation}

The formation  of the disk within  the halo leads to  a contraction of
the dark matter component. We  assume that the collapse of the baryons
within the dark halo is slow, and correct for adiabatic contraction of
the halo following  the procedure outlined in Barnes  \& White (1984),
Blumenthal  et al. (1986) and  Flores et al.  (1993). To  facilitate a
more direct  comparison with previous  studies, we do not  correct for
adiabatic  contraction   when  using  the   pseudo-isothermal  density
distribution.

\begin{figure*}[th]
\begin{center}
\epsfxsize=0.85\hsize \epsfbox{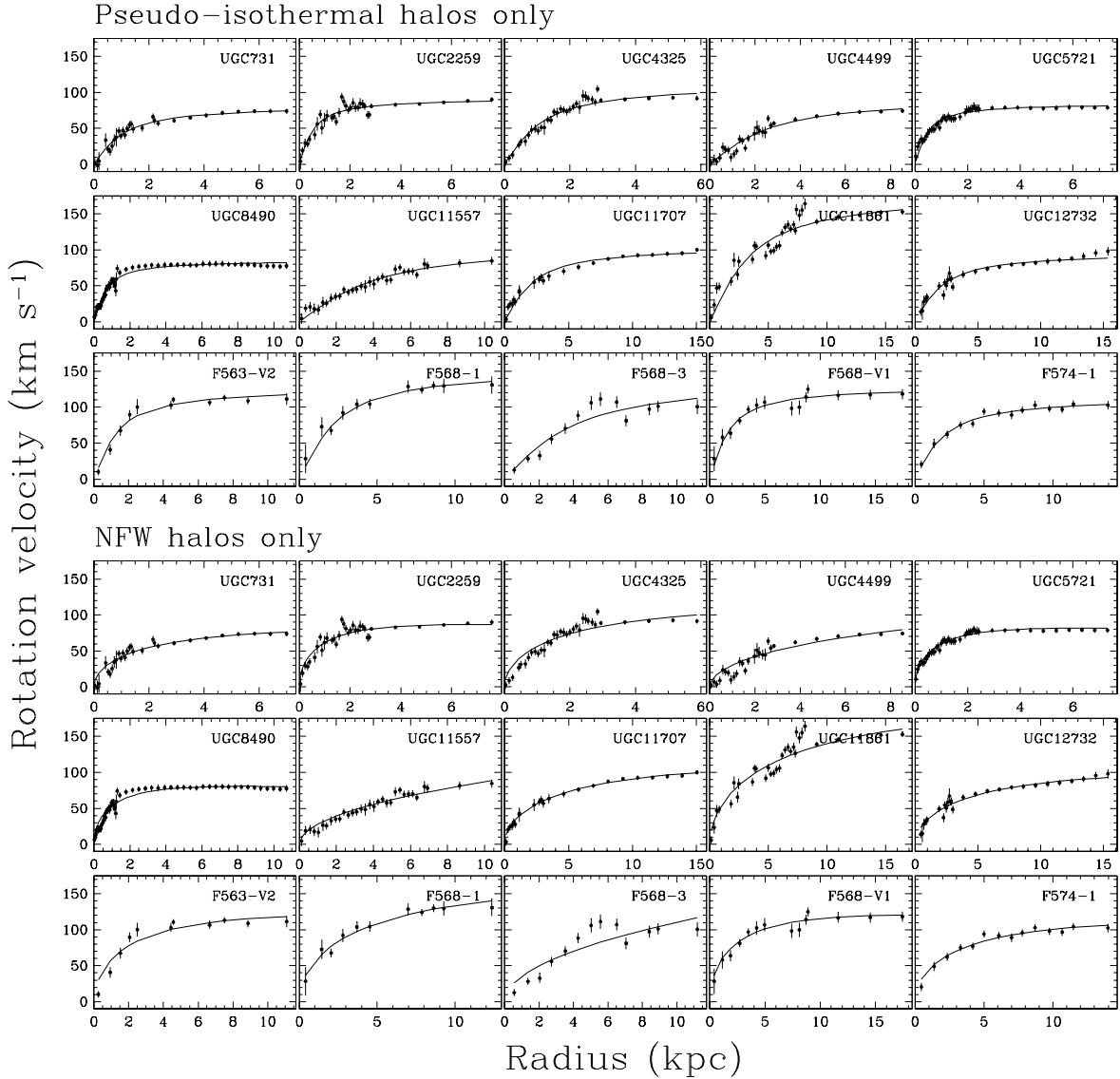}
\caption{Best fitting mass models for pseudo-isothermal halo (top) and
NFW (bottom). The filled circles represent the rotation curves as presented in
Fig.~\ref{figrcs}, the solid lines represent the best fitting dark
halos.  \label{figallfits}}
\end{center}
\end{figure*}

\section{Results}
\label{secresults}

Even when the effects of beam-smearing are small, a unique
decomposition of the observed RCs in terms of its contributions from
the stellar disk, the gaseous disk, and the dark matter halo is
hampered by uncertainties in the stellar mass-to-light ratio (the
mass-to-light ratio degeneracy, e.g., van Albada et al. 1985) and by
the limited spatial sampling of the the halo's density distribution
(the cusp-core degeneracy, vdBS).  Below we address this
non-uniqueness in the mass models by examining several extreme cases,
and by analyzing the data using different methods.

\subsection{Minimum disk models}
\label{secmindisk}

Useful limits on the dark halo properties and the slope of the dark
matter density distribution can be obtained by completely ignoring the
contribution of the disk (stars and gas), i.e., by assuming that the
dark matter is the only mass component.  These models are generally
referred to as `minimum disk' models.

\begin{table*}
\begin{center}
{\sc Table \ref{tabfitpars}\\ \smallskip\hbox to\hsize{\hfil{Best fit parameters}\hfil}}
\small
\setlength{\tabcolsep}{7.5pt}
\begin{tabular}{lccccccccccccc}
\tskip \tableline
\tableline \tskip
\colhead{} & \multicolumn{4}{c}{maximum disk and ISO} &
\multicolumn{3}{c}{ISO only} & \multicolumn{3}{c}{NFW only} & \colhead{} & 
\colhead{} & \colhead{} \\
\colhead{} & \multicolumn{4}{c}{\hrulefill} & 
\multicolumn{3}{c}{\hrulefill} & \multicolumn{3}{c}{\hrulefill} & \colhead{} & 
\colhead{} & \colhead{} \\
\colhead{Name} & \colhead{$\Upsilon_R$} & \colhead{$r_c$} &
\colhead{$\rho_0$} & \colhead{$\chi^2_\mathrm{r,max}$} &
\colhead{$r_c$} & \colhead{$\rho_0$} &
\colhead{$\chi^2_\mathrm{r,iso}$} & \colhead{$c$} & \colhead{$\varv_{200}$} &
\colhead{$\chi^2_\mathrm{r,nfw}$} & \colhead{$\alpha$} &
\colhead{$\Delta\alpha$} \\
\colhead{(1)} & \colhead{(2)} & \colhead{(3)} & \colhead{(4)} &
\colhead{(5)} & \colhead{(6)} & \colhead{(7)} & \colhead{(8)} &
\colhead{(9)} & \colhead{(10)} & \colhead{(11)} & \colhead{(12)} &
\colhead{(13)}\\
\tableline \tskip
UGC 731 & 15.1 & $\infty$ & 1.48 & 1.05 & 0.87 &  163 & 0.90 & 12.0 &   64 & 1.40 & 0.35 & 0.61 \\
UGC 2259 & 11.1 & $\infty$ & 4.34 & 3.93 & 0.45 &  751 & 1.78 & 23.9 &   57 & 2.27 & 0.86 & 0.18 \\
UGC 4325 & 8.95 & \nodata & \nodata & 1.16 & 0.94 &  263 & 1.46 & 14.8 &   83 & 3.43 & 0.26 & 0.33 \\
UGC 4499 & 1.39 & 2.79 & 17.0 & 2.22 & 2.08 & 37.4 & 1.47 &  3.9 &  130 & 3.35 & 0.47 & 0.23 \\
UGC 5721 & 2.95 & 1.05 &  123 & 2.22 & 0.39 &  874 & 2.62 & 24.8 &   53 & 1.54 & 1.16 & 0.17 \\
UGC 8490 & 2.25 & 1.13 & 95.6 & 0.92 & 0.56 &  424 & 1.00 & 19.4 &   56 & 1.66 & 0.83 & 0.16 \\
UGC 11557 & 1.09 & $\infty$ & 1.48 & 0.74 & 3.35 & 20.0 & 0.67 &  1.0 &  399 & 0.82 & 0.84 & 0.27 \\
UGC 11707 & 8.55 & 17.5 & 1.18 & 0.92 & 1.81 & 62.3 & 1.48 &  7.0 &   98 & 0.48 & 0.65 & 0.31 \\
UGC 11861 & 5.20 & 9.09 & 0.34 & 3.43 & 2.92 & 69.4 & 4.91 &  8.3 &  159 & 4.67 & -0.03 & 0.53 \\
UGC 12732 & 7.06 & 13.9 & 2.17 & 1.02 & 1.77 & 54.7 & 1.11 &  6.8 &   90 & 0.78 & -0.01 & 0.88 \\
F 563-v2 & 5.82 & \nodata & \nodata & 0.95 & 1.13 &  231 & 1.36 & 14.4 &   92 & 4.36 & 0.25 & 0.45 \\
F 568-1 & 15.8 & \nodata & \nodata & 1.71 & 1.95 &  115 & 0.50 & 10.3 &  129 & 0.62 & 0.67 & 0.12 \\
F 568-3 & 1.32 & 4.00 & 20.3 & 2.93 & 3.23 & 35.3 & 2.52 &  1.0 &  637 & 4.97 & 0.38 & 0.19 \\
F 568-v1 & 10.6 & 14.5 & 1.96 & 0.99 & 1.43 &  150 & 0.77 & 13.7 &   93 & 0.86 & 1.07 & 0.59 \\
F 574-1 & 4.56 & 9.44 & 0.17 & 0.95 & 1.66 & 86.3 & 0.54 &  9.6 &   93 & 0.99 & 0.78 & 0.34 \\
\tableline
\end{tabular}
\end{center}
\centerline{\vbox{\hsize=15.5cm \footnotesize 
Notes --- (1) the name of the galaxy. Best fit
parameters for maximum disk fits with a pseudo-isothermal halo are
listed in (2--5), for pseudo-isothermal halo only in (6--8), and for
NFW only in (9--11). (2) gives the mass-to-light ratio in
\Msun/\Lsun. (3) and (6) list the core radius in kpc, (4) and (7) the
central dark matter density in $10^{-3}$ \Msun\pc3. (9) gives the
central halo concentration index, (10) the halo velocity in \kms. (12)
and (13) give the inner slope measured from the mass density profile
and its error.}}
\end{table*}

In Fig.~\ref{figallfits} we show the best fitting minimum disk models
for both pseudo-isothermal and NFW dark matter halos. The best fitting
parameters are given in Table~\ref{tabfitpars}. Note that for the NFW
fits we impose $c\geq 1$ to prevent models in which the scale radius
$r_s$ is larger than the virial radius of the halo.  A visual
inspection of the fits shows that, in most cases, both the NFW and the
pseudo-isothermal halos appear able to describe the observed rotation
curves fairly well.  However, in a number of cases, the NFW model
overpredicts the inner slope of the rotation curve: UGC~731, UGC~4325,
UGC~4499, and F563-V2, and in two cases, UGC~5721 and UGC~11707, the
NFW model seems to fare better than the pseudo-isothermal model.  For
F568-3 and UGC~8490 both the pseudo-isothermal and the NFW give poor
fits to the observed rotation curves. Overall, the pseudo-isothermal
halo is somewhat more successful than the NFW halo in providing
well-fitting minimum disk models for the observed rotation curves, as
is also apparent from a comparison of the reduced \chisq\ values (see
Fig.~\ref{figchivschi}). This result is in agreement with findings by
other studies (Blais-Ouellette et al. 2001; dBMR; dBB).

If one assumes the disks are dynamically insignificant, and if one
assumes furthermore that the dark matter has a spherically symmetric
distribution, it is possible to recover the density distribution of
the dark matter from the observed rotation curve in a non-parametric
way.  From $\nabla^2\Phi=4\pi G\rho$ and $\Phi=-GM/r$ the density
distribution is given by
\begin{equation}
\rho(r) = {{1}\over{4\pi
    G}}\left(2{{\varv}\over{r}}{{\partial\varv}\over{\partial r}} +
    {{\varv^2}\over{r^2}} \right)
\label{eqinvert}
\end{equation}
This inversion method has been used by dBMBR and dBB to\break

\resizebox{0.99\hsize}{!}{\includegraphics{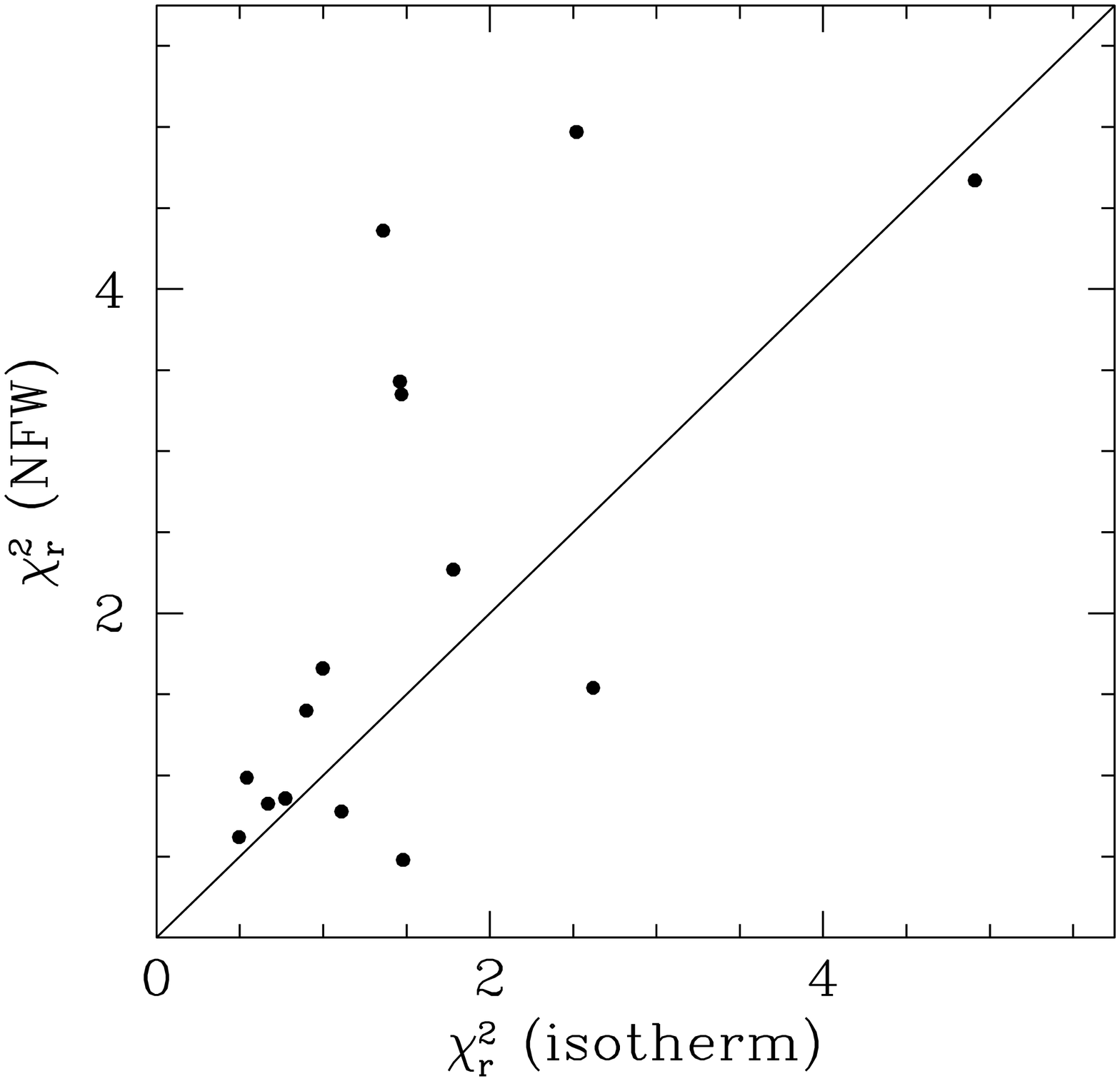}}
{\small {\sc Fig.~\ref{figchivschi}.}---
Comparison of the reduced $\chi^2$ values for the minimum
disk mass models with pseudo-isothermal and NFW halos. The solid line
is the line of equality.}
\bigskip
\addtocounter{figure}{1}

\noindent estimate the central cusp slopes of dark matter
halos. Following their procedures, we inverted the measured rotation
curve using Eq.~\ref{eqinvert}, and we identified by eye a break
radius inside of which we fit the density distribution with a simple
power-law: $\rho \propto r^{-\alpha_m}$.

\begin{figure*}[th]
\epsfxsize=1.00\hsize \epsfbox{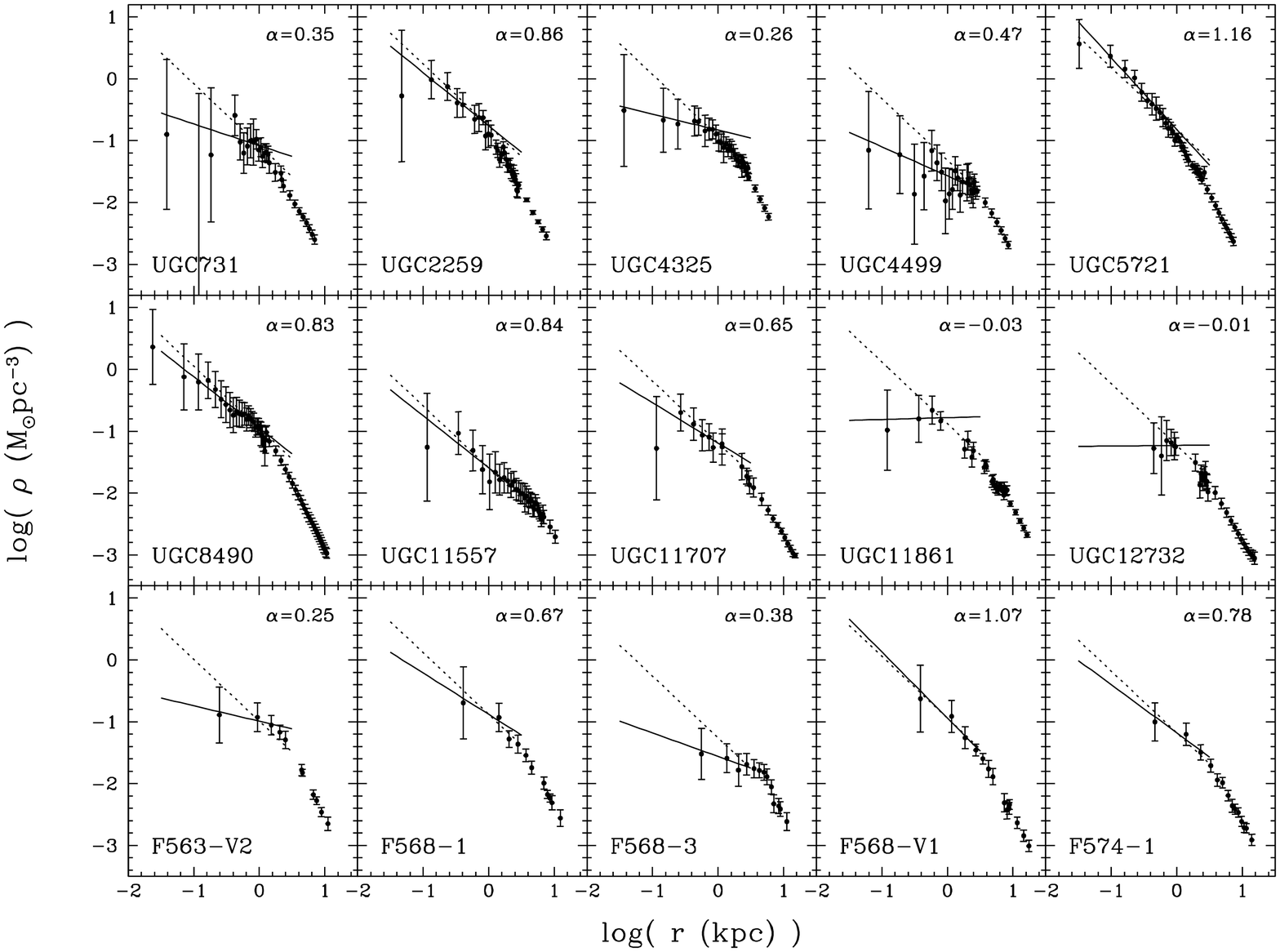}
\caption{ Log-log plots of the density distribution as derived by
inverting the observed rotation curves, following the procedure as
described in Section~\ref{secmindisk}. The solid line represents a
weighted least-squared fit to the parts of the profile within the
break radius, the dotted lines represent a slope of $\alpha=1$.
\label{figinvert}}
\end{figure*}

The derived density profiles, and the power-law slopes fit to their
inner parts are shown in Fig.~\ref{figinvert}. In
Table~\ref{tabfitpars} the inner slopes and their uncertainties are
listed. The uncertainties may be large, mostly because the derived
inner slopes depend on the derivative of the rotation curves, but also
because there may be little \Halpha\ emission in the center (e.g., UGC
12732), and because the derived inner slopes may depend on only a few
points with large error bars (e.g., F568-1).  The distribution of
measured inner slopes $\alpha_m$ is shown in Fig.~\ref{figdistralpha}.
Overplotted on this distribution is the corresponding relative
probability density distribution, derived by representing each entry
by a normalized Gaussian centered on its value with a width given by
its error, and coadding these.  The measured inner slopes span a wide
range with $0\la\alpha_m\la 1.2$, and are somewhat skewed toward
larger $\alpha_m$.

\subsection{Mass models with disk}
\label{secmaxdisk}

Although useful to obtain limits on the density distribution of the
dark matter, the minimum disk models presented above are most likely
not realistic. Therefore, we have also analyzed our rotation curves by
including the contribution of the baryons.

We followed the approach taken by vdBS and fit mass models consisting
of the generalized NFW profile of Eq.~\ref{eqrcgen}, the gaseous disk,
and the stellar disk.  For a given choice of $\mlstar$ and $\alpha$ we
determined the best fit values of $c$ and $\varv_{200}$, taking
adiabatic contraction into account, and not allowing values of $c$
less than unity.  The results are shown in Fig.~\ref{figgenfits}, in
which we plot $c$, $\varv_{200}$, and the reduced $\chi^2$ of the
best-fit models as function of $\alpha$ for four different values of
the stellar mass-to-light ratio $\mlstar = 0$, $0.5$, $1.0$, and
$2.0$, but only if those values are lower than the maximum disk
\mlstar\ as listed in Table~\ref{tabfitpars}.  This covers the typical
range of mass-to-light ratios expected from stellar population
modeling (e.g., Worthey 1994; Bell \& de Jong 2001).

A number of general trends are apparent.  The halo concentration $c$
of the best fitting model decreases with increasing $\alpha$, because
the enclosed mass for the different models has to be similar.  The
value of $\varv_{200}$ reaches a maximum where $c=1$.  For steeper
$\alpha$, the best fitting $c$ would be smaller than unity for which,
as mentioned above, Eq.~\ref{eqgennfw} no longer gives a meaningful
description of the dark matter distribution.  As a consequence, the
quality of the fits rapidly decreases. Finally, an increase of
\mlstar\ causes $c$ to decrease, because more and more of the inner
rise of the rotation curve is due to the stellar disk.  For most
galaxies the quality of the fit is similar for cusp slopes
in the range $0 \lta \alpha \lta 1.0$.  Only for four galaxies in our
sample do models with $\alpha=0$ yield significantly better fits than
for $\alpha=1$ independent of \mlstar\ (UGC~4325, UGC~4499, F563-V2 and
F568-3).  Finally, for $\alpha \gta 1.0$, the quality of the fit
decreases rapidly for virtually all galaxies.

These results are similar to those of vdBS, who fitted the same mass
models but to the HI rotation curves only.  For the galaxies in common
with vdBS (all UGC galaxies except UGC~2259, UGC~5721 and UGC~11557)
the results presented here are similar, suggesting that our higher
resolution \Halpha\ data have not strongly reduced the degeneracy in
the mass models.  Two exceptions are UGC~11861 and UGC~12732 for which
$\alpha$ has become better constrained with the new \Halpha\ data.

In summary, even with the high-resolution \Halpha\ data presented here
a unique decomposition of the rotation curves is not possible.  In
particular the freedom in the inner slope of the dark matter
distribution is large, and most galaxies {\it individually} are
consistent with mass models with inner slopes in the range
$0\la\alpha\la 1$. However, for $\alpha>1$ the quality of the fits
decrease rapidly, and inner slopes with $\alpha=1.5$ are clearly
inconsistent with the observations.
\vspace{1cm}

\section{Observational Effects}
\label{secobseffs}

Previous studies have found that the \Halpha\ and HI rotation curves
of dwarf and LSB galaxies are consistent with the picture that all
dark matter halos have constant density cores and that
pseudo-isothermal halos in general provide better fits to the rotation
curves than the cuspy NFW halos. The results presented in
Section~\ref{secresults} are consistent with these findings.  However,
there are various systematic effects that may affect the derived inner
slope, such as beam smearing, slit alignment errors, line-of-sight
projection effects and non-circular motions.  Clearly, such systematic
effects need to be considered before intrinsically steep slopes can be
ruled out observationally.

\subsection{Beam smearing}
\label{secbeam}

The effect of beam smearing is to dilute steep velocity gradients,
thus leading to an underestimate of the inner slope in the derived
rotation curves.  Previous attempts to constrain the central mass
density of LSB galaxies using HI rotation curves with a typical
resolution of $15''$ to $30''$ suffered strongly from beam smearing
effects. The \Halpha\ observations presented here have a much higher
spatial resolution, of about $1''$ to $2''$, and\break

\resizebox{0.99\hsize}{!}{\includegraphics{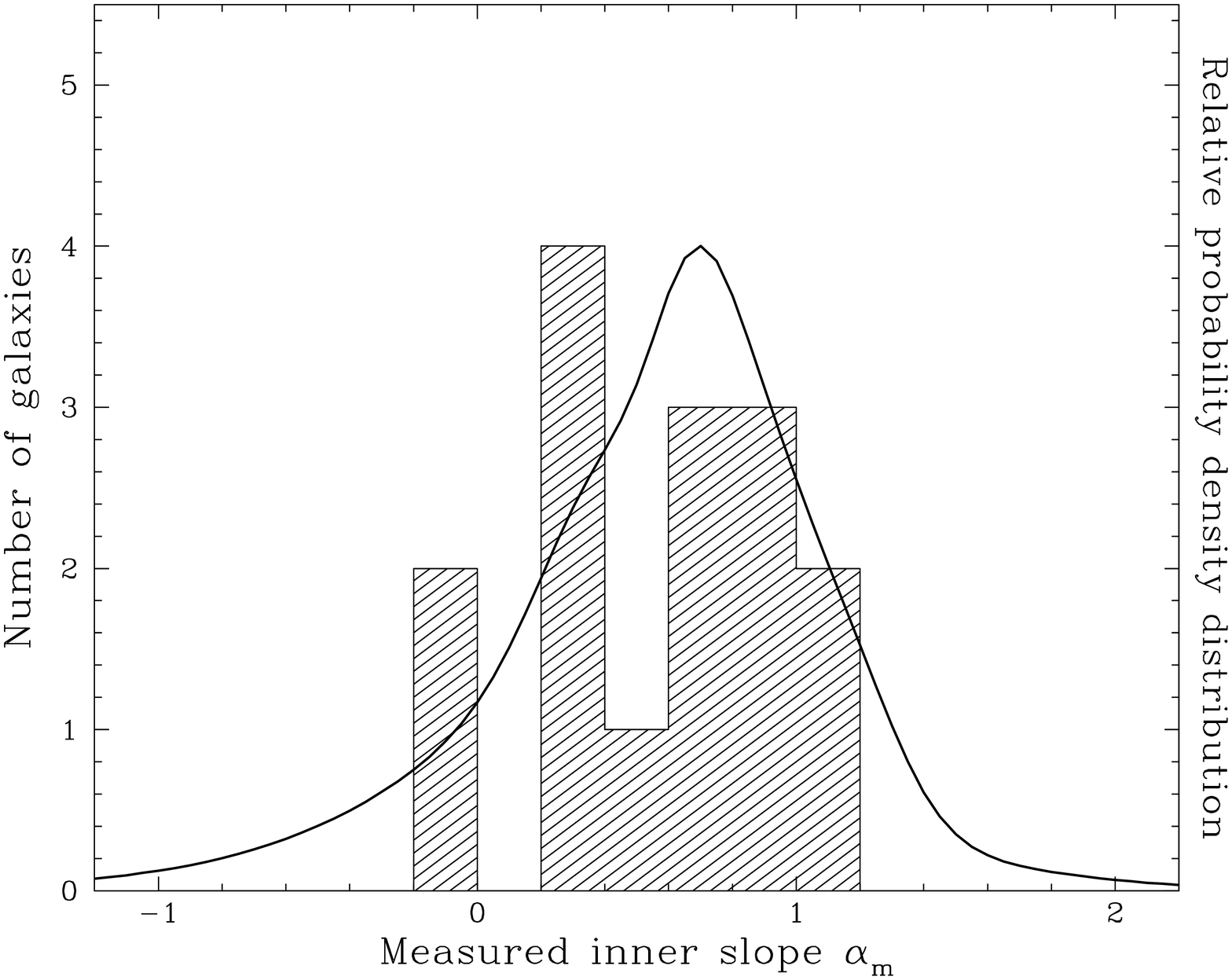}}
{\small {\sc Fig.~\ref{figdistralpha}.}---
Histogram of measured inner slopes $\alpha_m$. The solid line
is the corresponding relative probability density
distribution.}
\bigskip
\addtocounter{figure}{1}

\noindent clearly beam smearing will play a less dramatic role.
However, the measurement of the inner slope usually depends on the
very inner points of the rotation curves, and hence beam smearing as a
result of seeing still needs to be considered, especially for galaxies
at large distances.

\subsection{Slit width}

For a larger slit width, a larger fraction of the velocity field is
encompassed by the slit. The projected rotation velocities away from
the major axis are lower than those on the major axis, especially near
the center of the galaxy. This skews the observed line profile toward
lower rotation velocities, and hence a wide slit may lead to an
underestimate of the inner slope.

\subsection{Slit alignment errors}

A misalignment of the slit may happen either in position or in
position angle. The latter is likely to be negligible, both because
position angles can usually be measured accurately, and because even
large errors in position angle will still only have a small effect on
the rotation curve. A much more important effect is an offset between
the position of the slit and the center of the galaxy. As a result of
such an offset, the steepest gradient in the velocity field will be
missed, resulting in an underestimate of the inner slope.  There are
three main reasons for offsets: observational errors, uncertainties in
the position of the galaxy centers, and differences between the
kinematic and optical centers.

The observational errors depend on the method used to align the slit
with the galaxy image. Here, we have aligned the slit visually,
because most galaxies were clearly visible on the slit viewing
camera. For the LSB galaxies, we estimate that we could align the slit
with an accuracy of around $1''$. For the dwarf galaxies the alignment
was more difficult, because of their diffuse and extended nature, and
the error in the alignment will likely be larger.  A better method to
position the slit would be to use the coordinates of the optical
center, e.g., like dBB have done. They find that telescope control usually
allows a repeatable positioning with an accuracy of better than
$1''$. Their procedure reduces the observational uncertainty
significantly.

Another contributor to the alignment error is the accuracy of the
optical center itself. For many cataloged galaxies the optical center
has been determined from photographic plates. For the faint and often
irregular dwarf and LSB galaxies, the centers determined in this way
may well be incorrect by many arcseconds. But even if CCD photometry
with accurate astrometry is available, there may still be considerable
uncertainty in the position of the optical center. Swaters \& Balcells
(2002) have determined the centers of the 171 dwarf galaxies in their
sample, and published the centers of the isophotal fits versus
radius. Their figures show that there is considerable spread in the
best fitting centers as a function of radius, and that the centers of
the inner and outer isophotes usually differ by a few arcseconds, and
sometimes even as much as $30''$.

A third possible contributor to slit alignment errors is an offset
between the kinematical and optical centers. Such an offset is not
uncommon in dwarf and LSB galaxies (e.g., Puche, Carignan, \& Wainscoat
1991; Swaters et al. 1999). Obviously, if the dynamical and the
kinematical center do not agree, the steepest gradient in the velocity
field of the galaxy may be missed, and the inner slope underestimated.
Unfortunately, without two-dimensional velocity fields it is not
possible to assess whether an offset between kinematical and optical
center is present.

\begin{figure*}[ph]
\begin{center}
\epsfxsize=0.95\hsize \epsfbox{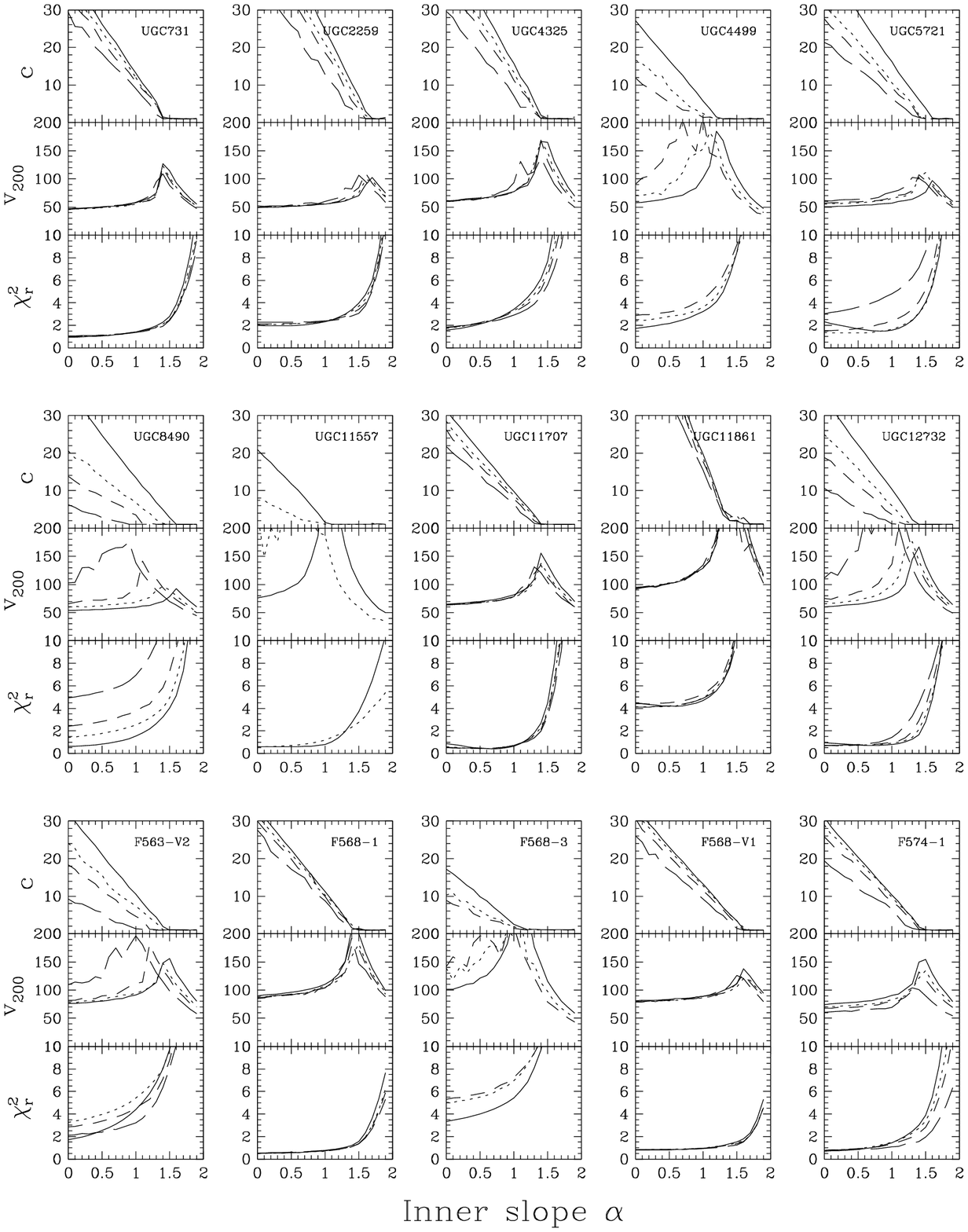}
\caption{Best fitting parameters and \rchisq\ for the generic halo
fits. The top panel for each galaxy shows $c$ versus $\alpha$, the
middle panel $\varv_{200}$, and the bottom one \rchisq. The different
lines represent fits with different \mlstar, where the solid line
represents $\mlstar=0$, the dotted line is $\mlstar=0.5$, the short
dashed lines is $\mlstar=1$ and the long dashed line is $\mlstar=2$.
\label{figgenfits}}
\end{center}
\end{figure*}

\subsection{Edge-on galaxies}
\label{secedgeon}

\begin{figure*}[th]
\begin{center}
\epsfxsize=1.00\hsize \epsfbox{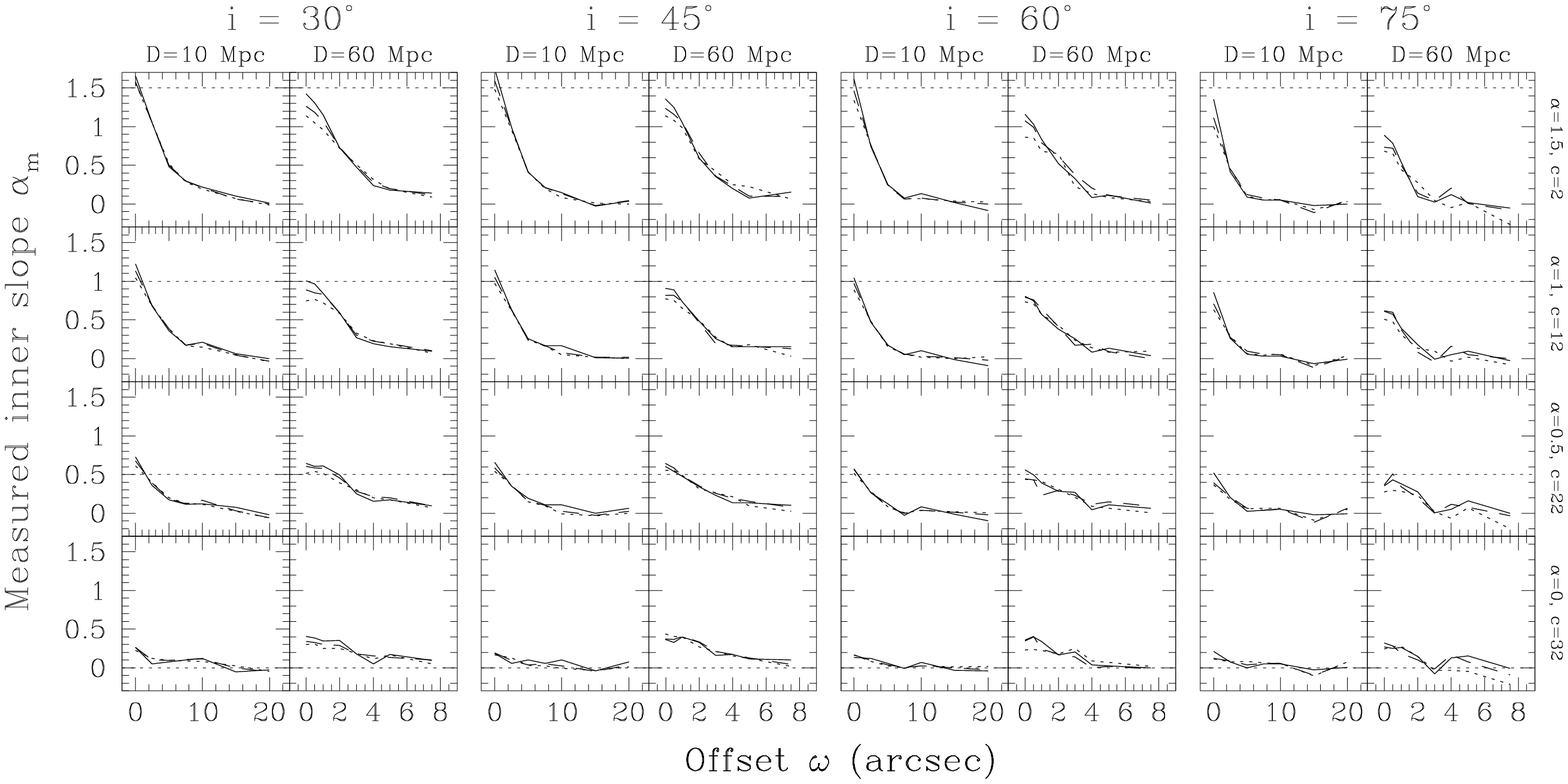}
\caption{The inner slope $\alpha$ as derived from the model
observations versus the offset of the slit from the major axis
$\omega$, for model galaxies at 10 Mpc and 60 Mpc, and a range of
inclinations as indicated above each column. The solid line represents
the results for a seeing of $1''$, the dashed line the results for
$1.5''$, and the dotted line the results for $2''$.The horizontal
dotted lines represent the intrinsic inner slope of each model.
\label{figmodalpha}}
\end{center}
\end{figure*}

In edge-on galaxies, the line-of-sight traverses the entire galaxy
disk, and hence a range of radii contribute to the observed velocity
profiles. However, only the gas on the major axis reflects the
circular velocity at that galactocentric radius, whereas gas from all
other radii has smaller projected velocities. The result is a broad
velocity profile that is strongly skewed toward lower rotation
velocities.  Simply taking the barycenter or a Gaussian fit will thus
result in an underestimate of the rotation velocity. To derive the
true rotation velocity, the extreme velocity should be used. To
determine this reliably a very high velocity resolution is needed,
much higher than is provided by the long-slit observations presented
here. In addition, determining the extreme velocity becomes more
difficult if the \Halpha\ emission is weak on the line of nodes.

\subsection{Distribution of \Halpha}

A more subtle cause of uncertainties in determining the dark matter
cusp slope is related to the spatial distribution of \Halpha\ in the
part of the galaxy that is visible through the slit.  If the \Halpha\
emission is uniform, this is not likely to play a role, but if the
\Halpha\ distribution is knotty, substantial underestimates may occur.
For example, consider the case in which there is no \Halpha\ emission
on the major axis, but that there is emission just next to it, but
still within the area viewed by the slit, or smeared into that area by
the seeing. The projected rotation velocity just next to the major
axis is lower than exactly on the major axis, and thus an irregular
distribution may lead to an underestimate of the derived rotation
curve. This effect is stronger when the seeing is poor, and it is
especially strong in the center, where small spatial differences may
lead to large velocity differences (see also Barth et al. 2001).  The
effects of the distribution of \Halpha\ can be modeled if an \Halpha\
image is available, allowing a more accurate measurement of the
rotation curve.

\subsection{Non-circular motions}
\label{secnoncirc}

Non-circular motions are another source of systematic errors in the
mass densities derived from rotation curves.  The usual assumption for
derived rotation curves is that the gas moves on circular orbits.
However, this assumption is not always accurate.  Structural features
such as bars, spiral arms, lopsidedness, the irregular distribution of
stars and gas, and the effects of star formation may all leave an
imprint on the kinematics.  The effects of such non-circular motions
are usually most important in the central regions, where rotation
velocities are relatively low and were the physical causes for
non-circular motions, such as bars and star formation activity, are
strongest.

It is difficult to model the effect of non-circular motions on the
derived rotation curve because many factors may contribute. For
example, a high star formation rate may lead to gas outflows and to an
increase the dispersion of the emission line gas, or of the gas
clouds, resulting in a lower azimuthal velocity component, and hence
an underestimate of the rotation velocity. The magnitude of this
asymmetric drift is uncertain and difficult to calculate, as was
pointed out in Section~\ref{secderrcs}.  The effects of bars on the
rotation curves depend on the bars' properties such as their
orientation and strength.  Other structural perturbations such as
lopsidedness or spiral structure, may either lead to an underestimate
or an overestimate of the rotation velocity, depending on the kind of
disturbance and its orientation.

\section{Modeling}
\label{secmodeling}

From Section~\ref{secobseffs} it is clear that there are strong
observational biases toward shallower slopes.  In this Section, we
derive more quantitative estimates of how these observational effects
affect the inner slopes derived from inversion of the rotation curves
and the mass modeling, based on model observations.  We focus on the
effects of the seeing, slit width, slit alignment errors and edge-on
galaxies, because their effects can be most readily determined from
model observations.

To construct the model velocity fields, we have assumed the model
galaxies to be infinitely thin disks that are uniformly filled with
\Halpha\ emission. We also experimented with alternative, more
complicated distributions, but found that these do not significantly
change the main results as long as the size of the seeing disk is
small compared to the typical length scale of variations in the
\Halpha\ emission. We assumed that the model galaxies are dominated by
their dark matter halos, and we considered dark matter halos with a
density distribution given by Eq.~\ref{eqgennfw}, with $\alpha$=0,
0.5, 1, and 1.5. The model with $\alpha=1$ has $c=12$ and $\varv_{200}=75$
\kms. For the other models, the values of $c$ and $\varv_{200}$ were
determined from a best fit to the $\alpha=1$ model, to ensure that the
rotation curves are similar in shape. Each model was placed at a
distances of 10 Mpc and 60 Mpc, and at inclinations of $30^\circ$,
$45^\circ$, $60^\circ$ and $75^\circ$.

After we constructed each model, it was convolved with the seeing and
the velocity dispersion. We made models for three different values of
the seeing, $1''$, $1.5''$ and $2''$, and assumed that the model
velocity dispersion is dominated by the instrumental one, as is also
the case for the galaxies in our sample.  Next, we extracted from the
model velocity field a spectrum for a given slit width and a given
offset from the model galaxy center, and derived a rotation curve
using the same procedure as for the real observations.  Because the
effects of seeing and slit width are similar, and because usually
observers match the slit width to the typical seeing, we have focused
only on models with equal slit width and seeing. Finally, we analyzed
these model rotation curves in the same way as the real data (see
Section~\ref{secresults}). The results are described in detail below.

\begin{figure*}
\begin{center}
\epsfxsize=1.00\hsize \epsfbox{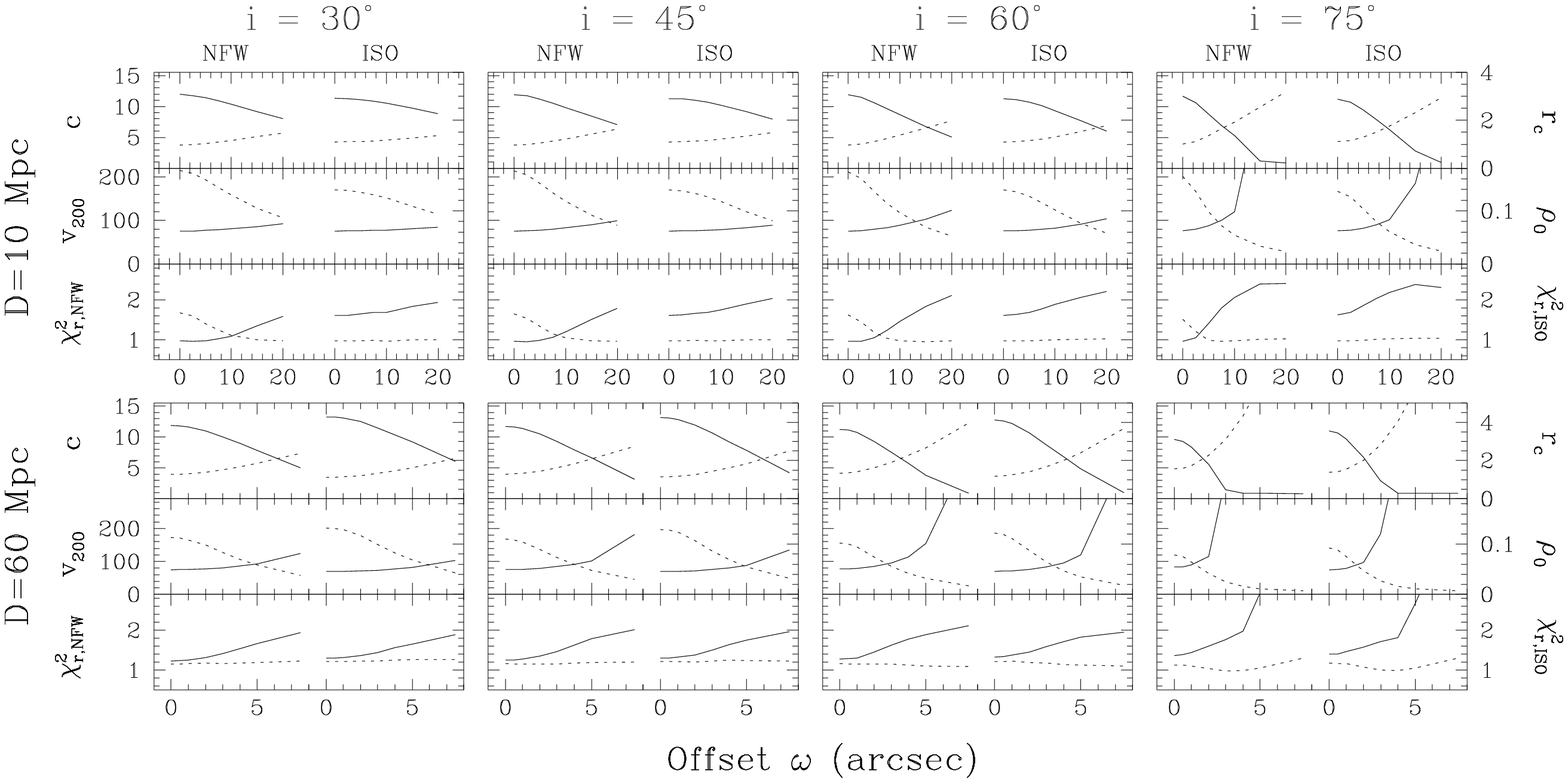}
\caption{Parameters for the best fitting mass models fitted to a model
with an NFW halo with $c=12$ and $\varv_{200}=75$ \kms\ (left column in each
panel) and a pseudo-isothermal halo with $r_c=1.1$ kpc and
$\rho_0=0.14$ \Msun\ \pc3\ (right column in each panel).  The solid
lines give the resulting parameters for an NFW fit, the dotted lines
for a pseudo-isothermal fit, both plotted versus the offset between
the slit and the major axis of the galaxy.  The results are virtually
independent of seeing, and hence we only show the model for a seeing
of $1''$. \label{figmodmassmod}}
\end{center}
\end{figure*}

\subsection{Inversion of the rotation curve}
\label{secmodinv}

We measured the inner slope by inverting the model rotation curves
using Eq.~\ref{eqinvert}, and following the same method as outlined in
Section~\ref{secmindisk}.  The results for model galaxies with
intrinsic inner slopes of $\alpha=0$, 0.5, 1 and 1.5 are shown in
Fig.~\ref{figmodalpha}, in which the measured inner slope $\alpha_m$
is plotted as function of slit offset $\omega$ for inclinations
$30^\circ$, $45^\circ$, $60^\circ$, and $75^\circ$, and for distances
of 10 and 60 Mpc.  The results are shown for models with a seeing of
$1''$ (full line), $1.5''$ (dashed line), and $2''$ (dotted line). The
horizontal dotted line in each panel represents the intrinsic inner
slope of each model.

If the slit is properly aligned with the major axis, i.e.,
$\omega=0''$, the inner slope can be accurately retrieved for the
galaxies at 10 Mpc, i.e., the effects of seeing and slit width are
negligible, provided that $i\la 60^\circ$. For galaxies with
inclinations around $75^\circ$, the inner slope is significantly
underestimated for the model galaxies with intrinsically steep slopes,
and this underestimate is larger for poorer seeing.

For the model galaxies at 60 Mpc, however, the effects of slit width
and seeing play a role at all inclinations. Even at low inclinations
and small seeing disks, the inner slopes may be significantly
underestimated. For example, for model galaxies with $\alpha=1$, the
inner slope is underestimated by around 20\% for $i=60^\circ$, and
around 40\% for $i=75^\circ$. Note that for galaxies with shallow
intrinsic slope the measured inner slope is actually an overestimate
of the intrinsic inner slope. As was also pointed out by dBMBR, for
these galaxies the innermost points already include the parts where
the density distribution turns over from its more shallow inner parts
to the steep $r^{-2}$ outer parts.

Fig.~\ref{figmodalpha} also shows that, if the slit is not aligned
correctly with the major axis, the measured inner slope may be
severely underestimated, both for $D=10$ Mpc and $D=60$ Mpc, for
galaxies with rotation curves with steep intrinsic slopes.  As the
offset increases, the velocity gradient over the slit becomes more
shallow, and for large offsets the measured inner slope may even
resemble a constant density core.  Note that the effect of slit
offsets does not scale linearly with distance.  For example, for the
model galaxy with $D=60$ Mpc, $i=45^\circ$, and $\alpha=1$, a measured
inner slope of $\alpha_m\approx 0.6$ is found for $\omega = 2''$.  One
might expect a similar value of $\alpha_m$ for $\omega = 12''$ for the
same model at $D=10$ Mpc, but in stead it occurs at about
$\omega=3''$.  This is a result of the sampling in combination with
the definition of a break radius.  For galaxies at smaller distances,
the inner parts of the rotation curves are well sampled, and the
effects of an offset become noticeable in the derived density
distributions even for small offsets. For galaxies at large distances,
the inner parts of the rotation curves are more poorly sampled, and
larger offsets are required to cause a noticeable change in the
density distribution.

In order to get a clear picture of the systematic effects as a result
of seeing, slit width and offsets, we have not included realistic
errors on the rotation curves.  The scatter seen in
Fig.~\ref{figmodalpha} is the result of discretisation noise in the
models. We have repeated the modeling and added Gaussian noise to the
rotation velocities with an amplitude of $5 \kms$.  This results in a
typical error on the measured inner slope $\alpha_m$ of 0.25.

\subsection{Mass model fitting}
\label{secmodmassmod}

We have fitted mass models to the rotation curves derived from the
models for different values of the offset $\omega$, seeing, and slit
width, in order to investigate how accurately the halo parameters can
be retrieved.  In these fits, the intrinsic inner slope was kept fixed
because for realistic errors on the derived rotation velocities the
cusp-core degeneracy prevents an accurate determination of the inner
slope, as is clear from the general dark matter fits presented in
Fig.~\ref{figgenfits} and from the models presented in vdBS.  We
focused on the NFW halo (which has $\alpha=1$) and the
pseudo-isothermal halo (which has $\alpha=0$), since these are most
often used in the literature.

\hskip-0.3cm
\resizebox{0.99\hsize}{!}{\includegraphics{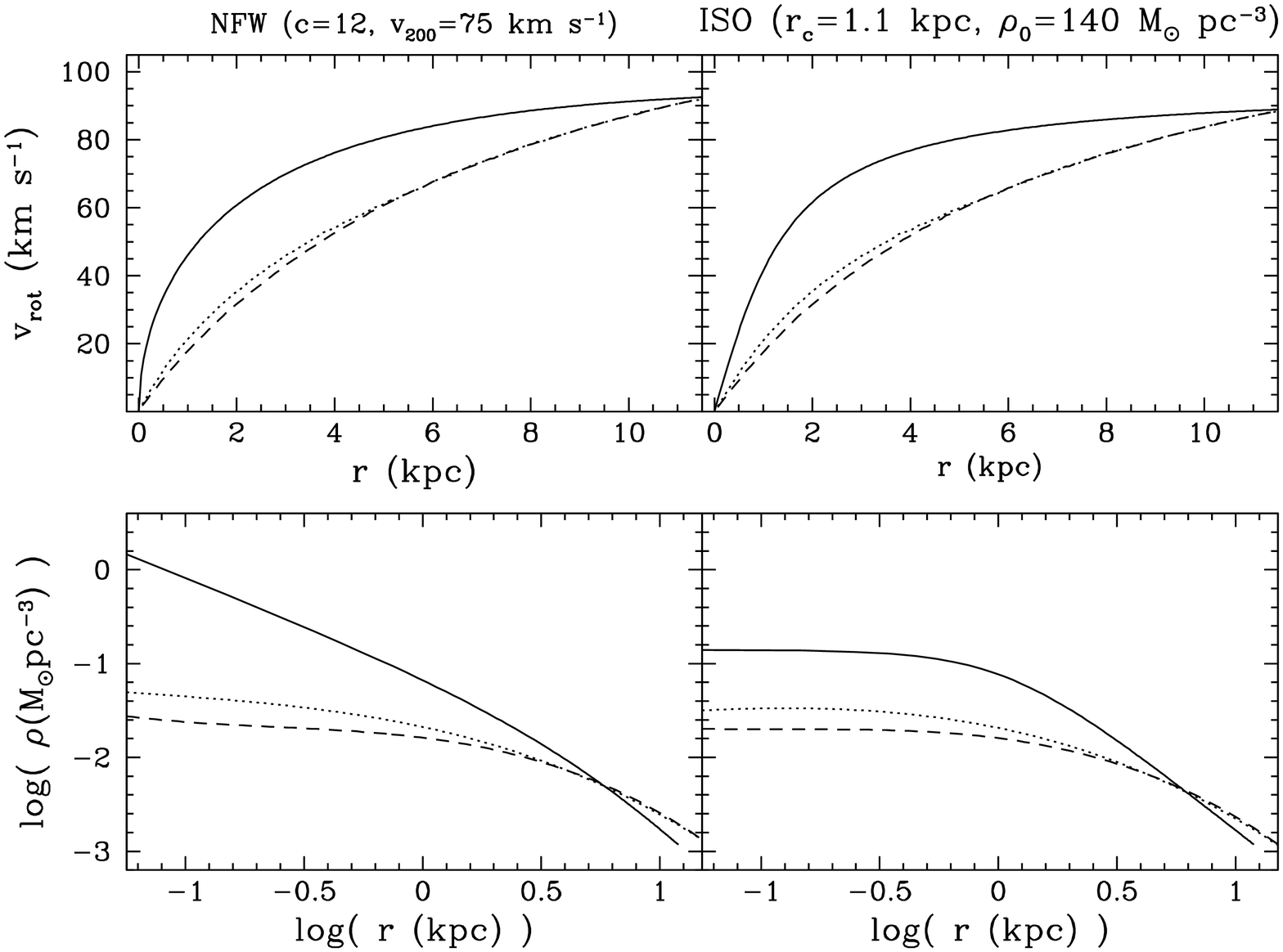}}
{\small {\sc Fig.~\ref{figmodedgeon}.}---
The top panels show the input rotation curve ({\it solid
lines}) and the rotation curves as derived from the model observations
of an edge-on galaxy with $1''$ seeing, for a constant density
distribution ({\it dashed lines\/}) and an exponentially declining
distribution ({\it dotted lines\/}). The bottom panels show the input
and the derived density distribution, with the line coding the same as
in the top panel.  The left column gives the results for an NFW halo,
the right model for a pseudo-isothermal halo. The halo parameters are
given above each column.}
\bigskip\bigskip
\addtocounter{figure}{1}

We assumed that the model galaxies are dominated by either an NFW halo
with $c=12$ and $\varv_{200}=75$ \kms, or a pseudo-isothermal halo with
$r_c=1.1$ kpc and $\rho_0=110$ \Msun\pc3.  We fit both
pseudo-isothermal and NFW halos to both sets of models, for the same
range in seeing, slit width, offsets and inclination used in
Section~\ref{secmodinv}, and for distances of 10 Mpc and 60 Mpc.  For
all galaxies, we considered the model rotation curves out to a maximum
radius of 10 kpc.  To mimic the properties of the observed rotation
curves as closely as possible, we have resampled the model rotation
curves in the same way as the observed ones. For the model galaxies at
10 Mpc, the rotation curves were sampled every $2''$ in the inner 3
kpc, and every $15''$ at larger radii. This closely resembles the
combination of \Halpha\ and \HI\ data for the closeby galaxies. The
model galaxies at 60 Mpc were sampled every $2''$ out to the last
measured point.  We have added Gaussian noise to the rotation curves
with $\sigma=5 \kms$.

In Fig.~\ref{figmodmassmod} we present the results. In each panel, the
left column shows the fits to the models built from an NFW halo, and
the right column shows the fits built from a pseudo-isothermal
halo. Each model has been fit both with an NFW halo ({\it full
line\/}) and a pseudo-isothermal sphere ({\it dotted line\/}). As can
be seen in Fig.~\ref{figmodmassmod}, the input parameters can be
retrieved fairly accurately if the slit is aligned with the major
axis, and if the rotation curves are fit with the correct model. We
found that the effects of seeing and slit width are generally
negligible, and therefore we only show the results for a seeing of
$1''$ in Fig.~\ref{figmodmassmod}. Seeing and slit width start to play
a minor role in the model galaxies at 60 Mpc and with inclination
$i\ga 60^\circ$, in which case errors in the derived parameters of up
to about 10\% may occur.

If the slit is not correctly aligned with the major axis, much larger
errors are introduced. At 60 Mpc, an offset of a few arcseconds is
sufficient to lead to a significant error in the derived parameters,
especially for highly inclined systems. At 10 Mpc, the impact of slit
offsets is much less pronounced, and the derived parameters are hardly
affected even for offsets as large as $5''$, except in highly inclined
systems, where the effects become significant for smaller offsets.
Overall, the results derived from mass modeling are less sensitive to
the effects of seeing, slit width, and slit offsets than the results
from rotation curve inversion. This is expected, because the entire
rotation curve is used to fit the mass models, whereas only the inner
parts of the rotation curve, which are most affected by seeing, slit
width, and slit offset, are used to determine the inner slope.

To investigate the usefulness of $\chi^2$ as a discriminator between
the two mass models, we have calculated and plotted the $\chi^2$
values for both the NFW and the pseudo-isothermal halos. For galaxies
at small distances, the $\chi^2$ appears to be a useful
discriminator. If model galaxies with pseudo-isothermal halos are fit
with NFW halos, the fits are poor. The same is true for the converse,
although for large slit offsets the pseudo-isothermal fits to the NFW
models may lead to the incorrect conclusion that these models have
pseudo-isothermal halos. The required offsets are large and
unrealistic, except perhaps in galaxies with high inclinations. At a
distance of 60 Mpc, the value of $\chi^2$ is not a useful
discriminator. Even if the slit is well aligned with the major axis,
the $\chi^2$ values for both models are comparable. If the slit is
somewhat offset from the major axis, the pseudo-isothermal fits
consistently result in lower $\chi^2$ than the NFW models, even if the
intrinsic halo shape is NFW. To be able to reliably distinguish
between the models presented here, errors on the rotation velocities
of about 1 or 2 \kms\ are needed. Given the often weak \Halpha\
emission, and the likely presence of small and large-scale
non-circular motions, it seems unlikely that the circular velocity can
be derived with sufficient accuracy.

\subsection{Edge-on galaxies}
\label{secmodedgeon}

As mentioned in Section~\ref{secedgeon}, projection effects may
strongly affect the derived rotation curves for galaxies that are seen
close to edge-on.  Here, we show two models to illustrate the
systematic effects.

Because the line of sight traverses the entire disk in edge-on
galaxies, the results will depend somewhat on the adopted radial
distribution of the \Halpha\ emission, and we therefore constructed
two sets of models, one with constant level of \Halpha\ emission, and
one with a radially exponential decline. As before, we assumed the
galaxies are dominated by their dark matter halos, and we used the
same halo models as in Section~\ref{secmodmassmod}. The model galaxies
were placed at a distance of 10 Mpc, at in inclination of $90^\circ$,
and ``observed'' with a $1''$ slit and $1''$ seeing. Finally, the
rotation curves are derived using the same procedure as used for the
real observations, i.e., by fitting Gaussians to the line profiles. We
have also used the barycenter to derive the rotation curve, and this
gives virtually identical results.

In Fig.~\ref{figmodedgeon} we compare the input model rotation curves
(solid lines) to the ones derived from the edge-on model galaxies. In
addition, we compare the input density distributions with those
obtained from the inversion of the model rotation curves, using
Eq.~\ref{eqinvert}. Dashed (dotted) lines correspond to models with a
constant (exponentially declining) level of \Halpha\ emission. From
Fig.~\ref{figmodedgeon} it is clear that for edge-on galaxies the
rotation curves derived from long-slit observations using traditional
methods suffer from extremely strong systematic effects. Hence, the
dark halo parameters derived from these rotation curves are likely to
be inaccurate.

\newpage

\section{Discussion}
\label{secdisc}

\subsection{Limits on the inner slopes}
\label{secdiscinn}

Our results presented in Section~\ref{secresults}, based both on a
measurement of the inner slope by inversion of the observed rotation
curve and on fitting mass models, have shown that the measured inner
slopes derived for the galaxies in our sample span a wide range from
$\alpha_m\sim 0$ to $\alpha_m\sim 1.2$. Taken at face value, the fact
a substantial fraction of the galaxies in our sample have a measured
inner slope around zero may lead to the conclusion that these galaxies
cannot have cuspy inner slopes. However, as discussed in
Section~\ref{secobseffs}, there are many systematic effects that may
affect the measured inner slopes, and most of these will lead to an
underestimate.  We will now use the modeling results from
Section~\ref{secmodeling} to try and determine which range of
intrinsic inner slopes is consistent with the range in measured inner
slopes. We will only consider the range $0<\alpha\la 1$. Slopes
steeper than $\alpha=1.5$ are ruled our by the observations presented
here, and by several previous studies (e.g., dBMBR; vdBS).

\subsubsection{Inner slopes from mass density profiles}

In Fig.~\ref{figdistralphall} we show the distributions of measured
inner slopes for the galaxies in our sample with $v_\mathrm{sys}\le
2000$ \kms\ (which have distances more typical of the model galaxies
at 10 Mpc, presented in Section~\ref{secmodeling}) and with
$v_\mathrm{sys}>2000$ \kms\ (which have distances more typical of the
model galaxies at 60 Mpc).  We will refer to these subsamples as the
10 Mpc and the 60 Mpc sample. Overplotted on the
histograms are the expected distributions for different intrinsic
inner slopes of the dark matter distribution, for different values for
the error in the slit position, and a slit with of $1''$ (the slit
width that we used in our observations).  The expected distributions
have been calculated as follows. From our modeling results presented
in Fig.~\ref{figmodalpha}, we know for each model halo with intrinsic
inner slope $\alpha$ what the measured inner slope $\alpha_m$ will be
for a given offset. Assuming a Gaussian distribution for the slit
positioning error $\sigma_\mathrm{off}$, we can also calculate the
probability that a particular $\alpha_m$ will be measured.  While
calculating this probability distribution we assumed that $\alpha_m$
can be measured with no observational error. However, in our
observations we found a typical error on the measured inner slope of
0.25 (see Table~\ref{tabfitpars}). To compare the expected
distribution of $\alpha_m$ with the observed one, we have therefore
convolved the expected one with a Gaussian with
$\sigma_m=0.25$. Finally, we selected for each galaxy the model with
the inclination closest to the inclination of that galaxy, and we
averaged the probability distributions for all galaxies in our sample
to produce the expected distributions shown in
Fig.~\ref{figdistralphall}.

For each adopted halo model we show the expected distributions for
different values of the offset error.  It is uncertain what the actual
amplitude of the slit offset error is, given that telescope pointing
errors, uncertain positions of the galaxy centers, and possible
offsets between the morphological and dynamical center all may play a
role. Therefore, we plot several expected distributions for different
values of the offset error $\sigma_\mathrm{off}$.  The solid lines
give the distributions of measured inner slopes expected if galaxies
are dominated by halos that have inner slopes $\alpha=1$ and offset
errors of $0''$, $1''$, and $2''$, as indicated above each curve. The
distributions expected if galaxies are dominated by halos with inner
slopes $\alpha=0$ are almost insensitive to the offset errors,
because any offsets will result in a measured inner slope close to
zero (see bottom row in Fig.~\ref{figmodalpha}). Hence, we only show
one curve for the $\alpha=0$ case.\penalty-10000

\vspace*{-0.7cm}\hspace{-0.8cm}
\resizebox{\hsize}{!}{\includegraphics{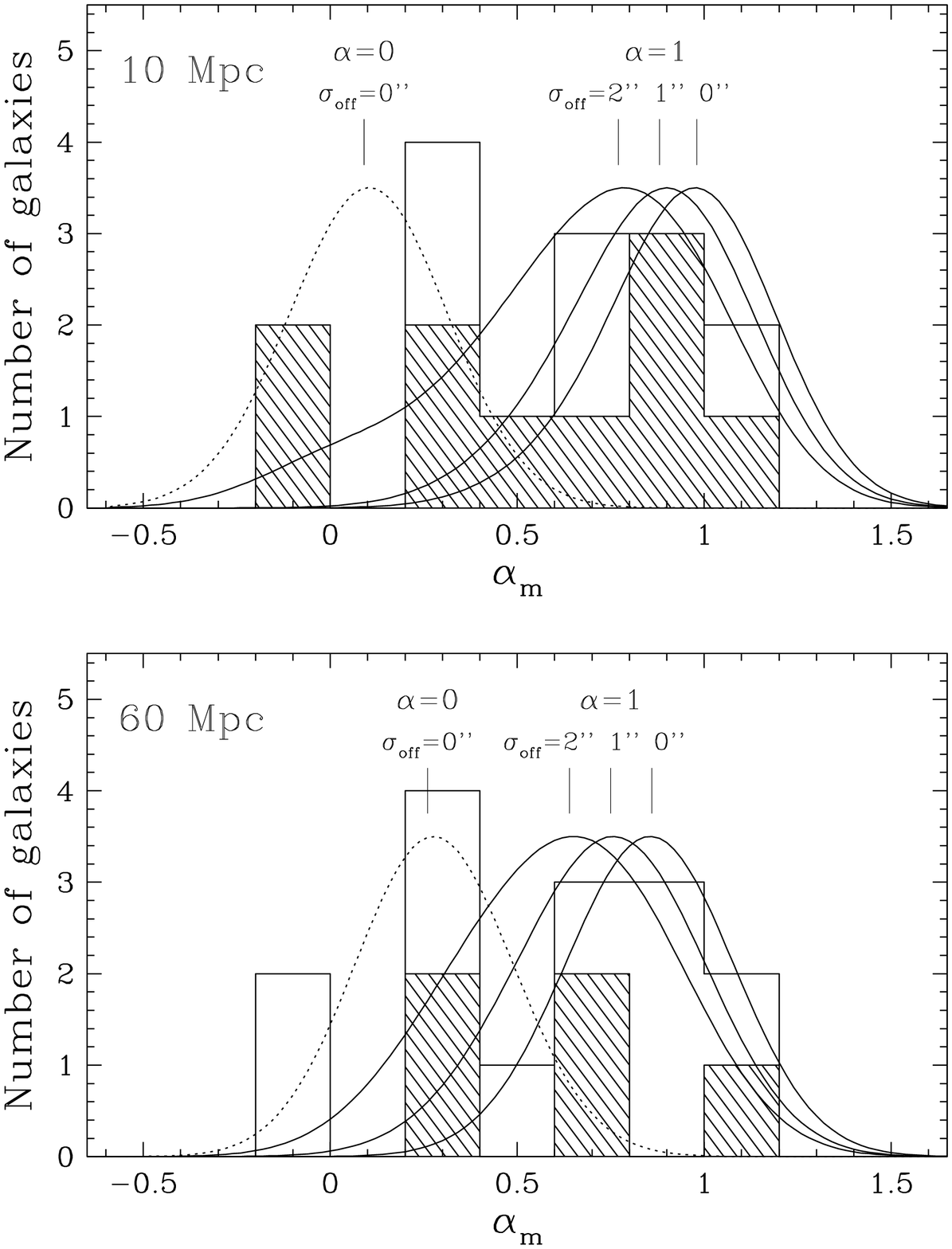}}
{\small {\sc Fig.~\ref{figdistralphall}.}--- 
The unshaded histograms in both panels represent the
distribution of inner slopes derived from fits to the central parts of
the density profiles as given in Fig.~\ref{figdistralpha}. The shaded
area indicates the galaxies in the distance group as indicated in the
top left of each panel. The bell-shaped curves represent the expected
distribution of measured inner slopes for the adopted intrinsic inner
slope $\alpha$ and slit alignment error $\sigma_\mathrm{off}$ as
labeled above each curve. A slit width of $1''$, a seeing of $1''$ and
a measurement error on $\alpha_m$ of 0.25 was assumed. How these
curves were derived is explained in the text.}
\bigskip\bigskip

A comparison of the expected distribution with the observed one is
difficult, because of the small number of galaxies, especially in the
60 Mpc sample. Still, it seems that the expected distribution for
$\alpha=0$ appears inconsistent with the observed one, both in the 10
Mpc and 60 Mpc samples. In both cases, there appear to be too many
galaxies with high $\alpha_m$.  The expected distribution based on
models with $\alpha=1$ is in reasonable agreement with the number of
galaxies with steep measured inner slopes, but it underpredicts the
observed number of galaxies with shallow inner slopes in the 10 Mpc
sample if $\sigma_\mathrm{off}$ is small. About 25\% of the galaxies
in our sample appear to have $\alpha_m$ inconsistent with $\alpha=1$,
unless the alignment error is larger.  If $\sigma_\mathrm{off}$ is
about $2''$, one expects to observe a wide range in $\alpha_m$, even
as low as $\alpha_m=0$, and the expected distribution is fairly
consistent with the observed one.

It appears that the observed distribution of measured inner slopes in
the 10 Mpc sample is not uniquely consistent with any of the models
overplotted. This may simply be the result of the small number of
galaxies in our sample.  Alternatively, it could be that one or more
of our assumptions does not hold, e.g., non-circular motions may play
a role, the mass distribution may not be spherical, or the stellar
disk is not massless.

\null\vskip-3pt
\hskip-0.3cm
\resizebox{0.99\hsize}{!}{\includegraphics{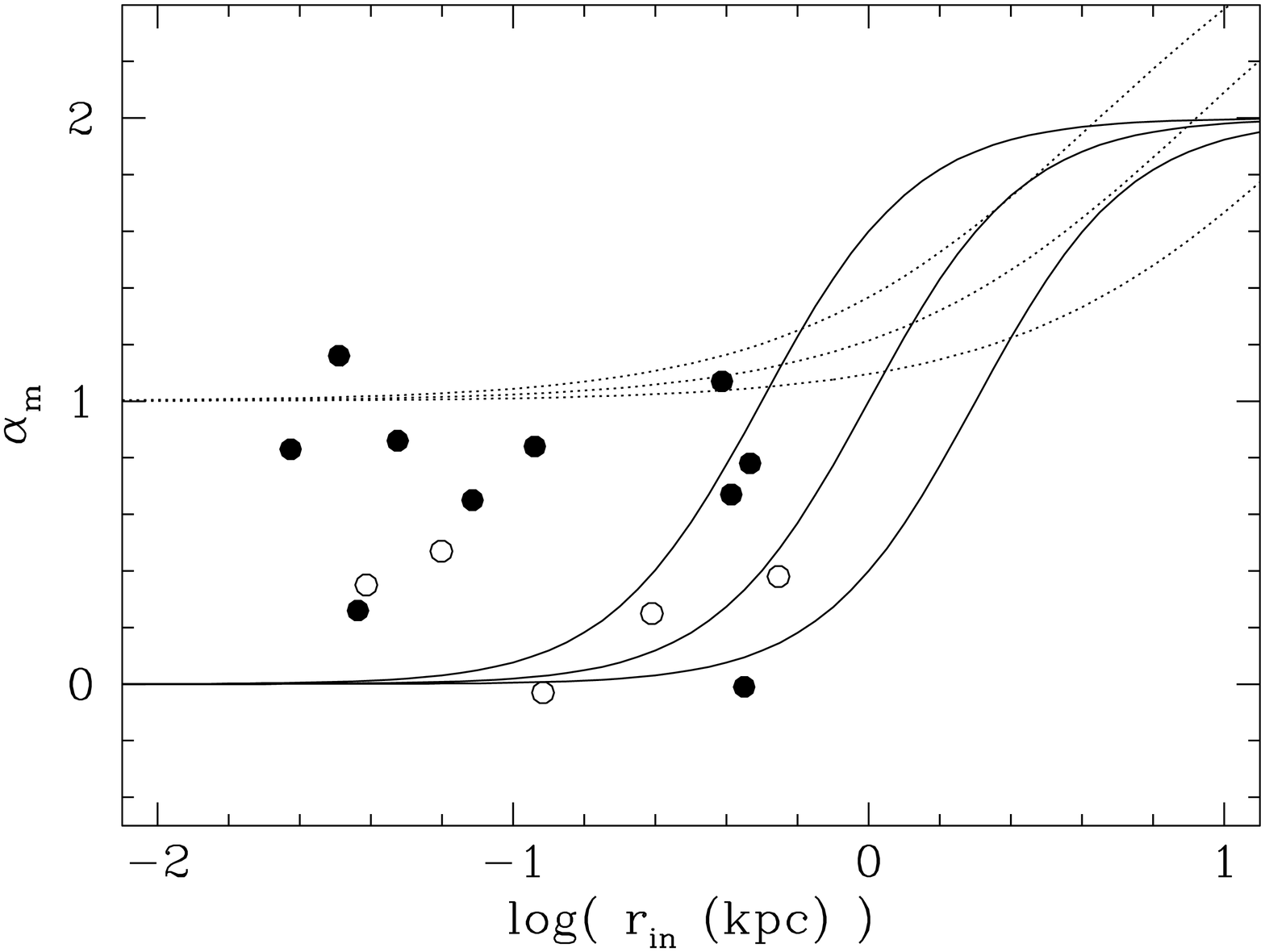}}
{\small {\sc Fig.~\ref{figalpharin}.}---
Measured inner slope $\alpha_m$ versus the radius of the
innermost point in the rotation curve. The solid lines are the inner
slopes of pseudo-isothermal halos with core radii, from left to right,
of $r_c=0.5, 1, 2$ kpc. The dotted lines are the inner slopes for NFW
halos with, from left to right, $c=18$ and $\varv_{200}=60$ \kms,
$c=12$ and $\varv_{200}=75$ \kms, and $c=6$ and $\varv_{200}=90$ \kms.
The filled circles represent non-barred galaxies, the open circles
represent barred galaxies.
}\bigskip

Of course, stellar disks do have mass, and the more dynamically
significant they are, the shallower the inner slopes of the dark halos
would have to be.  In this light, it is interesting to note that the
steepest inner slopes are measured in the galaxies with the highest
surface brightnesses and presumably the most massive disks (UGC~5721
and UGC~8490), suggesting that for these galaxies the contributions of
the stellar disks may have resulted in steeper measured inner
slope. On the other hand, we also found galaxies with steep inner
slopes among galaxies with low surface brightnesses (e.g., UGC~2259
and F568-V1).  These galaxies are dominated by dark matter for stellar
mass-to-light ratios of up to a few, and hence the measured inner
slopes likely reflect the inner slopes of the dark halos.

The inner slope as derived from the mass density profile has as
advantage that it is a parameter free measurement and that it directly
measures the inner slope of the halo. However, because it is usually
measured from only the innermost points, it is especially sensitive to
systematic effects. Also, $\alpha_m$ generally has a large error,
as it is derived from a density profile that depends on the
derivative of the observed rotation curve.  Because of these
uncertainties and systematic effects that may have affected the data,
our conclusion is that inner slopes in the range $0<\alpha<1$ are
consistent with the observations presented here.

A point of concern raised by dBMBR is that the steep inner slopes do
not represent the slopes of the inner parts of the dark matter
profiles, but rather the $r^{-2}$ outer parts. If this were the case,
one would expect to see shallow slopes in the best resolved galaxies,
and the steepest slopes in the galaxies that are the least
well-resolved. To verify this, we used the same test as dBMBR and
plotted $\alpha_m$ against the radius of the innermost point on the
rotation curve in Fig.~\ref{figalpharin}. Also drawn are the
inner slopes as a function of radius for different
pseudo-isothermal halos and NFW halos. From Fig.~\ref{figalpharin} it
is clear that we do find steep inner slopes among the best resolved
galaxies, and that the steep slopes of the galaxies in our sample are
not the result of the inclusion of the $r^{-2}$ outer parts.

\subsubsection{Inner slopes from the mass modeling}
\label{secdiscmassmod}

Our model observations presented in Section~\ref{secmodmassmod} have
shown that the parameters derived from the mass modeling are less
sensitive to the effects of seeing, slit width and slit offsets,
because the entire rotation curve is used to determine the parameters,
and not only the innermost part. As a consequence, the mass modeling is
also less sensitive to the particular value of the inner slope of the
dark matter halo (the cusp-core degeneracy), and for most galaxies we
find that mass models based on $\alpha=1$ provide fits of similar
quality as those based on $\alpha=0$. This is not only true if the
contribution of the stellar disk is ignored, but also for models in which
the stellar disks play a modest role.

Nonetheless, in about 25\% of the galaxies the mass models with
$\alpha=1$ give significantly poorer fits than those based on
$\alpha=0$, even if the baryonic component is ignored (UGC~4325,
UGC~4499, F563-V2 and F568-3).  Even though our modeling in
Section~\ref{secmodmassmod} has shown that the $\chi^2$ value may be
biased by slit offsets, unrealistically large offsets would be
required to make the observed rotation curves consistent with a halo
with $\alpha=1$. Hence, it is not likely that the poor fits for these
four galaxies are the result of alignment errors, except perhaps in
the case of UGC~4325, in which the optical center and the dynamical
center derived by S99 are offset by $5''$.

A possibility is that some of these poor fits are the result of
non-circular motions. It is difficult to get an estimate of the
strength of non-circular motion from the observations presented here,
as our long-slit observations only provide a one-dimensional slice
through the velocity field. On the other hand, the optical morphology
may provide an indication of whether non-circular motions play a role.
As a first order indicator of the presence of non-circular motions, we
used the presence of a bar.  The galaxies in our sample with either
mild or strong bars are UGC~731, UGC~4499, UGC~11861, F563-V2, and
F568-3.  Interestingly, 3 out of the 4 galaxies which are poorly fit
by an NFW halo have strong bars (UGC~4499, F563-V2 and F568-3),
suggesting that the non-circular motions associated with bars may have
contributed to the poor fits.

In conclusion, we find that almost all galaxies in our sample
are consistent with pseudo-isothermal halos. About three quarters of
our sample is also consistent with NFW halos, and the remaining
quarter may have been affected by non-circular motions.  Based on the
sample of galaxies presented here, and given the uncertainties, we
find that slopes in the range $0\la\alpha\la 1$ are consistent with
the observations.

\subsection{Comparison to the literature}

Our conclusion that the inner slopes of the dark halos seem consistent
both with $\alpha=0$ and $\alpha=1$ appears to be add odds with the
conclusions from dBMBR and dBB, who found that the derived inner
slopes for the galaxies in their sample peaks at $\alpha_m=0.2$, with
a tail toward steeper slopes.

Here we investigate whether the systematic effects studied in this
paper may have played a role in their work.  To this end, we have
divided the galaxies from the dBMBR and dBB sample in three
categories, based on a visual inspection of their optical images:
edge-on galaxies, barred galaxies, and the remaining galaxies.  We
have separated the galaxies into two distance groups, one with
galaxies that are closer than $2000 \kms$ , and one with galaxies that
are more distant than that.  As before, the separation in these two
distance groups was done to match\break

\vspace*{-0.5cm}
\hskip-0.3cm\resizebox{\hsize}{!}{\includegraphics{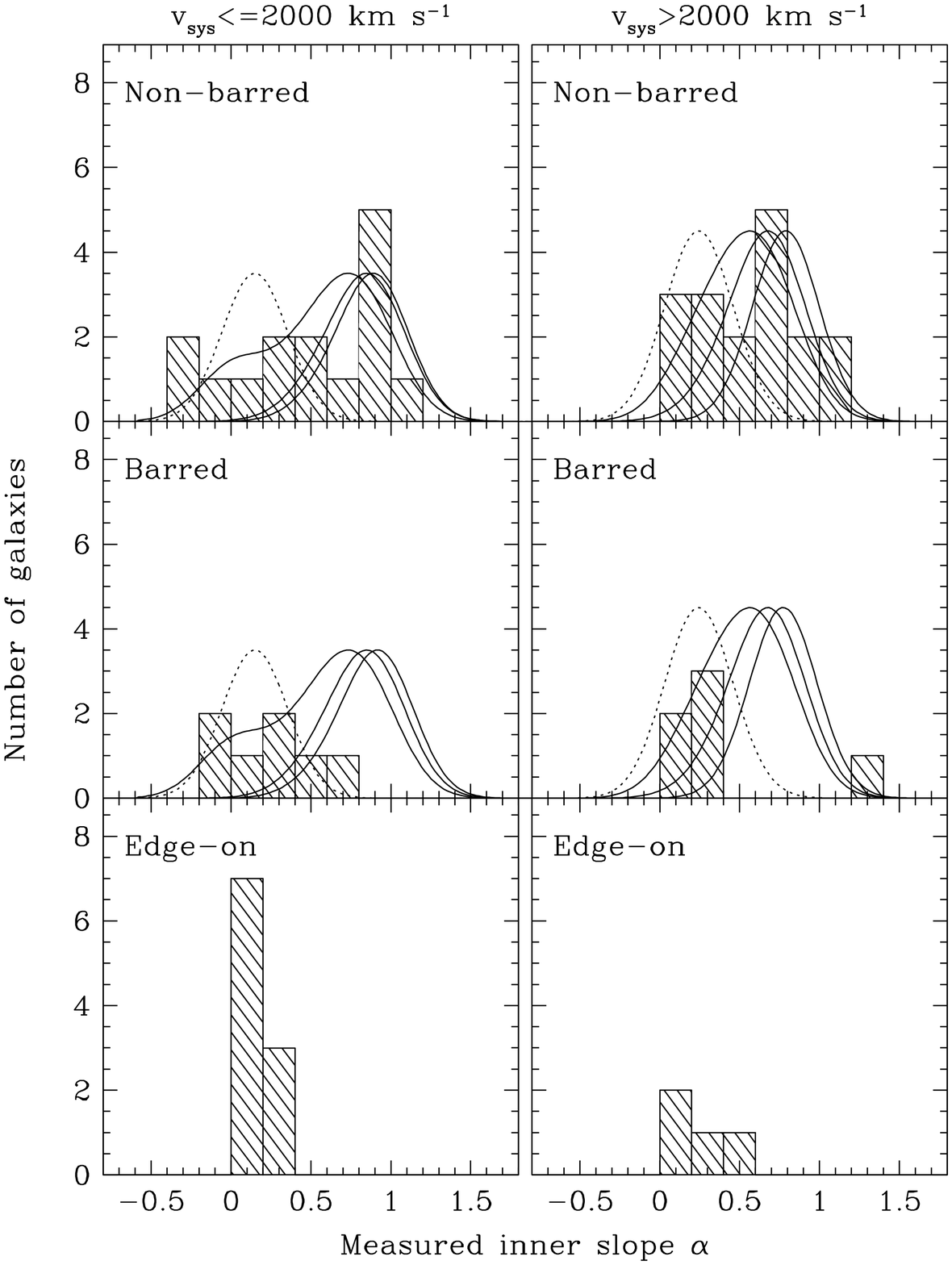}}
{\small {\sc Fig.~\ref{figalphalit}.}---
The distribution of measured inner slopes for the galaxies
in the samples of de Blok et al. (2001a), de Blok \& Bosma (2002), and
the sample presented here, divided into three groups, both for
galaxies within $2000 \kms$ (left column), and for galaxies beyond
$2000 \kms$.  The shaded histograms in the top panels show the
distribution of slopes for non-barred galaxies, the ones in the middle
panels for galaxies that are barred, and the ones in the bottom panels
for edge-on galaxies.  The dotted lines represent the expected
distributions of the inner slopes for $\alpha=0$, the solid lines from
left to right give the expected distributions for $\alpha=1$, with a
slit alignment $\sigma_\mathrm{off}$ of $2''$, $1''$ and $0''$. A
measurement error on $\alpha_m$ of 0.25 was assumed.
}\bigskip\bigskip

\noindent the separation in these two distance groups was done to
match the distances used in the models presented in
Section~\ref{secmodeling}.  We have excluded the three LSB galaxies
that dBMBR took from the Verheijen (1997) sample, because these were
based on low-resolution \HI\ observations.

We present the distributions of measured inner slopes for each of the
categories in Fig.~\ref{figalphalit}.  The galaxies in the sample
discussed in this paper are also included. For galaxies that are
present in both our and the literature studies, we have used the
results presented here.  Using the modeling results from
Section~\ref{secobseffs}, we calculated the expected distribution of
observed inner slopes for the non-edge-on and non-barred galaxies,
given an intrinsic inner slope of the dark matter distribution, a slit
with of $1.5''$ (the average of the observations in the different
samples) and different values for the slit alignment error, for
galaxies within $2000 \kms$, and for galaxies beyond $2000 \kms$.  The
expected distributions have been calculated in the same way as
described in Section~\ref{secdiscinn}, and are overplotted on the
histograms in Fig.~\ref{figalphalit}.\penalty-10000

\vspace*{-0.1cm}
\hskip-0.3cm\resizebox{\hsize}{!}{\includegraphics{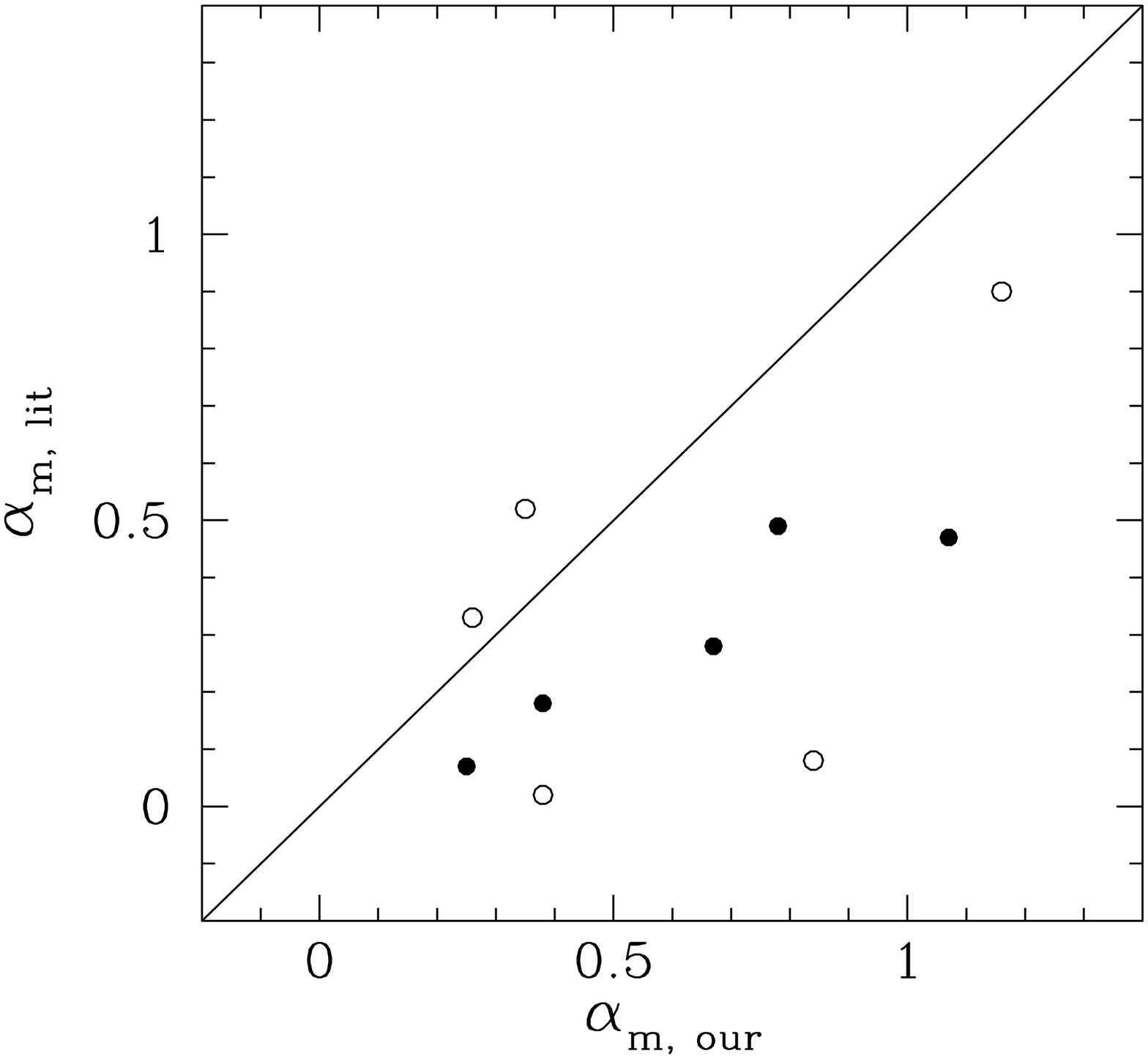}}
{\small {\sc Fig.~\ref{figcompalpha}.}---
Comparison of the values of the inner slopes as derived from density
profiles presented in this paper to those in the literature. The
filled circles represent galaxies for which the same data was used
both here and de Blok et al. (2001a), the open circles represent
galaxies for which independent observations were used.
}\bigskip\bigskip

For the non-barred galaxies with $v_\mathrm{sys}<2000$ \kms, the
measured inner slopes span a wide range from $\alpha=0$ to $\alpha=1$,
and for the closeby galaxies some galaxies even appear to have
negative $\alpha$, i.e., a dark matter density that increases with
radius.  The distribution of $\alpha_m$ for the non-barred galaxies in
the combined sample appears inconsistent with the distribution
expected for halos with $\alpha=0$, because of the excess of galaxies
with steeper inner slopes.  However, it is possible that the stellar
disks do contribute significantly and this may explain the wing to
high $\alpha_m$ even if the halos have $\alpha=0$.  The models with
$\alpha=1$ appear to underpredict the number of galaxies with
$\alpha_m$ near zero, unless the error in the slit positioning is
about $2''$, for which the model is good agreement with the observed
distribution. An error of $2''$ in the slit positioning may not be
unreasonable given the large angular size of these galaxies and their
irregular, diffuse and low surface brightness nature, and the fact
that pointing errors, errors in their optical centers, and possible
offsets between optical and dynamical center all contribute.

For the galaxies at 60 Mpc the expected distributions for the different
intrinsic halo shapes are more similar, although again it appears as
if the model for $\alpha=0$ appears to underpredict the number of
galaxies with steep slopes.

As before, we have used the presence of a bar as an indicator of the
presence of non-circular motions. The distribution of $\alpha_m$ for
the barred galaxies is shown separately in Fig.~\ref{figalphalit}.
Although based on only a few galaxies, the barred galaxies on average
appear to have a somewhat lower $\alpha_m$.  This may be an indication
that the non-circular motions associated with the bar have affected
the inner slopes.  On the other hand, if these bars were randomly
oriented, one would expect that such non-circular motions would mostly
lead to an increased scatter in the rotation curves, as observed by
e.g., Sofue et al. (1999), and hence on $\alpha_m$, but hardly to
systematically lower values.  However, in most of the galaxies in both
our sample and the literature sample, the bars are not randomly
oriented, but preferentially aligned with the major axis in almost all
galaxies. This might explain the observed shallower gradients in these
galaxies. Alternatively, the barred galaxies in our sample may have
halos with intrinsically shallow inner slopes.

A striking feature in Fig.~\ref{figalphalit} is the large difference
in the distribution of measured inner slopes between edge-on galaxies
and the non-barred galaxies, in both distance groups. Although perhaps
a result of the possibility that galaxies selected to have low surface
brightnesses when seen edge-on have different intrinsic properties
than those selected to have low surface brightnesses when seen
face-on, a more likely explanation is that this marked difference is
artificial.  As demonstrated in Section~\ref{secmodedgeon}, the method
used here and in several literature studies is not suited to derive
rotation curves for edge-on galaxies, as it results in a severe
underestimate of the inner slope.  In fact, as shown in
Section~\ref{secmodedgeon}, this method, when applied to edge-on
galaxies dominated by NFW halos, yields an inner slope of
$\alpha_m\sim 0.2$, close to the peak dBMBR found.

The sample presented in this paper has a number of galaxies in common
with the samples presented in dBMR and dBB. In Fig.~\ref{figcompalpha}
we plot the inner slopes as derived in the literature version those
derived there. The filed circles represent the inner slopes derived
for the five LSB galaxies presented in SMT. For these five galaxies
{\it the same\/} data have been analyzed in different ways by dBMR and
by us.  The open circles represent galaxies for which dBB
and we have independent observations.  As is clear from
Fig.~\ref{figcompalpha}, the inner slopes derived in this paper
generally are steeper than those derived in the literature. The
reasons for this difference are not clear.  Probably the choice of the
break radius within which the inner slope is measured, the fact that
dBMR and dBB have derived their rotation curves by
fitting smooth curves to their data, and the fact that the data have
been binned in $2''$ intervals here and in dBMBR, but $6''$ for some
galaxies in dBB all play a role.  Irrespective of
the cause for the difference, it is clear that when different
observers find such different results for the same galaxies, even when
the same data are used, the data do not allow firm conclusions to be
drawn.

In addition, we compared the derived values for $c$ and $\varv_{200}$
for the galaxies that we have in common with dBMR and dBB. In contrast
with the results for the derived inner slopes, the derived halo
parameters show little or no systematic difference between our and
their best fit values, although there is a considerable scatter of
about 35\% between the best fit values for $c$.  The fact that there
is a systematic difference between the derived inner slopes but hardly
in the derived halo parameters is not unexpected.  The inner slope is
determined solely from the inner few points. On the other hand, the
mass modeling uses the entire rotation curve, and the outer parts of
the rotation curves both in our and in the literature samples are
similar in shape because they both use the \HI\ data from S99.

We have also compared our fitting results to those obtained by 
Marchesini et al. (2002), who have three galaxies in common with the 
our sample (UGC~4325, UGC~4499, and UGC~11861). The fits presented in 
their Fig.~5 strongly overpredict the inner rise of the rotation curves 
for their NFW models. However, as Marchesini et al. (2002) warn, they 
did not fit the observed rotation curves, but only normalized each 
curve to its last point. When instead they make fits using the same 
mass models as described in Section~\ref{secmassmodels}, they find best 
fit values for $c$ and $v_{200}$ in good agreement with our results.
\vspace{0.5cm}

\hskip-0.3cm\resizebox{\hsize}{!}{\includegraphics{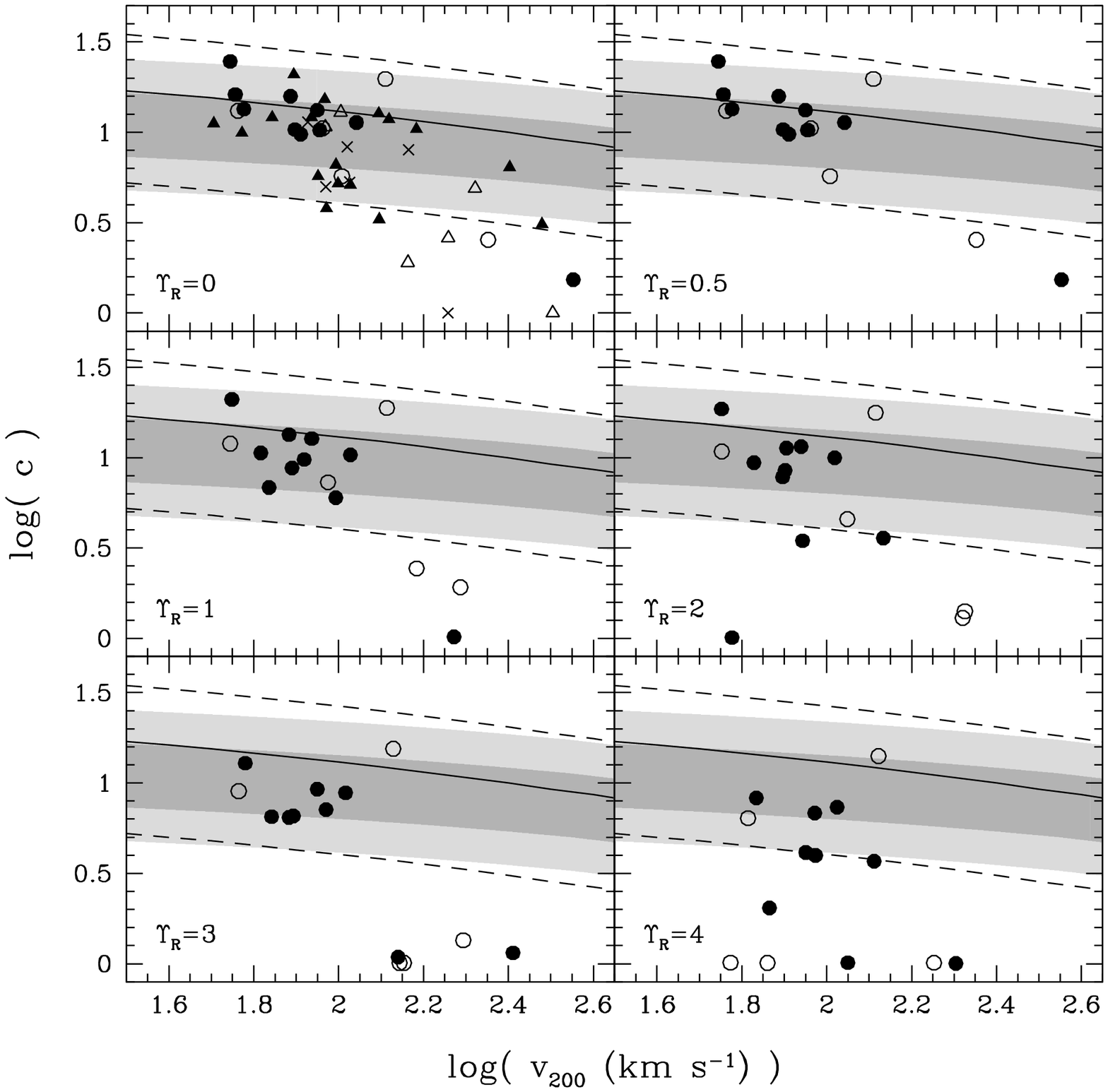}}
{\small {\sc Fig.~\ref{fignfwvsml}.}---
The NFW halo concentration $c$ plotted against the halo
rotation velocity $\varv_{200}$ for \mlstar\ of 0 to 4, as indicated in
the lower left corned of each panel. The filled circles represent the
non-barred galaxies in our sample, the open circles represent the
barred galaxies in our sample. The filled triangles are the non-barred
galaxies in the combined samples of de Blok et al. (2001b) and de Blok
\& Bosma (2002), the open triangles are the barred galaxies, and the
crosses the edge-on galaxies. The dark and light shaded regions
represent the $1\sigma$ and $2\sigma$ limits on the distribution of
halo parameters as predicted by Navarro et al. (1997), and the solid
and dashed lines represent the mean and $2\sigma$ limits from the
Bullock et al. (2001) model.}\bigskip\bigskip

\subsection{The nature of dark matter}
\label{secdisccdm}

The properties of dark matter halos depend on the nature of dark
matter and the cosmological parameters.  Here we compare the
predictions made from simulations based on the currently popular
$\Lambda$CDM paradigm.  There is disagreement between the different
cosmological simulations as to what the expected inner slope of the
dark matter halos in a universe dominated by CDM is. As noted in the
introduction, several studies have reported an inner slope with
$\alpha=1.5$, but the most recent simulations seem to prefer
$\alpha\sim 1$. This difference is crucial, as none of the galaxies in
our sample or the literature sample are consistent with $\alpha=1.5$,
whereas most of them, given the involved uncertainties, seem
consistent with $\alpha=1$.  At the same time, none of these galaxies
{\it require\/} halos with $\alpha=1$, as their rotation curves are usually
equally well or better explained by halos with shallower density
profiles or even constant density cores.

In Fig.~\ref{fignfwvsml} we compare the best-fit values of $c$ and
$\varv_{200}$ obtained from fitting mass models to the observed rotation
curves with the predictions from the numerical simulations of Navarro
et al. (1997) and Bullock et al. (2001). The open and filled circles
represent the halo parameters found for the barred and the non-barred
galaxies in our sample, the open and filled triangles are the barred
and the non-barred galaxies in dBMR and dBB, and the crosses are the
edge-on galaxies in their sample. We have plotted the results for a
range in M/L for the galaxies in our sample. Note that the fitting
results are slightly different from the results listed in
Table~\ref{tabfitpars}, because in the fitting here the contribution
of the \HI\ has been taken into account.

Fig.~\ref{fignfwvsml} shows that the majority of the galaxies have
values for $c$ and $\varv_{200}$ that are consistent with
$\Lambda$CDM, provided the M/L of the stellar disk is about 2 or
smaller.  M/Ls in this range are consistent with predictions from
stellar population synthesis models. Based on the typical $B-R$ color
of $0.84\pm 0.14$ for the galaxies in this sample (de Blok et
al. 1995, Swaters \& Balcells 2002), M/Ls in the range 0.4 to 1.4 are
expected (e.g., Bell \& de Jong 2001). It should be noted, however,
that these synthesis models do not provide a normalization of the M/L,
and changes in the adopted properties of the stellar initial mass
function will change the expected M/Ls.  If the M/Ls are larger than
2, more and more galaxies become inconsistent with the predictions of
$\Lambda$CDM. One galaxy from our sample, UGC~11557, as is also
apparent from Table~\ref{tabfitpars}, has a low $c$ as derived from
the best fit. However, as is clear from Fig.~\ref{figdata}, the
rotation curve for this galaxy does not extend much beyond the
turnover radius, and hence $c$ is poorly constrained. Even though
$c=1$ is the best fit for this galaxy, larger values, up to $c=4$ are
also consistent with the observed rotation curve.

Barred galaxies seem to have somewhat lower values for $c$, and, as
shown in Figs.~\ref{figdistralphall}, \ref{figalpharin}
and~\ref{figalphalit} they also tend to have shallower inner
slopes. Although this may be due to non-circular motions, it may also
indicate that the barred galaxies in our sample have halos with
intrinsically shallower inner slopes. Similar results have been found
from observations of barred spiral galaxies. The results from Weiner
et al. (2001) for NGC~4123 and Binney \& Evans (2001) for the Galaxy
indicate that the disks in these galaxies are close to maximal,
resulting in low density halos with shallow cores. Such shallow cores
appear to be in conflict with the predictions by $\Lambda$CDM
simulations, although one might envision a scenario in which the
interaction between a bar and a cuspy dark halo results in the
formation of a constant density core (Weinberg \& Katz 2002).

\section{Summary and conclusions}
\label{secconcl}

We presented high resolution \Halpha\ rotation curves for a sample of
15 dwarf and LSB galaxies derived from long-slit observations.  From
these, we measured the slope of the central mass distribution, both
from the mass density distribution implied by the rotation curves, and
from fitting mass models to the rotation curves. 

To assess the influence of systematic effects on the derived inner
slopes, we have modeled the effects of galaxy inclination, slit width,
seeing and slit alignment errors on the derived rotation curves.
Other systematic effects that are less easily modeled, such as
non-circular motions or the detailed spatial distribution of \Halpha\
emission, have been discussed but not modeled.  Combined, these
systematic effects may lead to a systematic bias toward shallower
slopes if the intrinsic halo profile has a steep inner slope. If the
inner slope is intrinsically shallow, the inner slope is hardly
affected, and may even be somewhat overestimated.

Using the data presented in the paper, and recent data for similar
galaxies presented in the literature, we find that both the inner
slopes as derived from the mass density profiles and from the mass
model fitting indicate that most galaxies in our sample are consistent
with the range in intrinsic inner slopes $0\la\alpha\la 1$. Inner
slopes with $\alpha>1$ are ruled out by these observations.  The range
of intrinsic inner slopes that appear consistent with the observations
is rather large, because the inner slopes as derived from the mass
density profiles are susceptible to the systematic effects mentioned
in this paper, and because the mass modeling is not very sensitive to
the inner slopes (the cusp-core degeneracy).  We note that about 25\%
of the galaxies in our sample appear to be inconsistent with NFW
halos. These galaxies are predominantly barred, suggesting either the
rotation curves for these galaxies may have been affected by the
non-circular motions associated with the bars, or perhaps that the
halos of these galaxies intrinsically have shallower slopes.  The
remaining 75\% have best-fit values for their halo parameters that are
consistent with the values expected in a $\Lambda$CDM cosmology,
provided that the $R$-band stellar mass-to-light ratios are smaller
than about 2.

Even though the majority of the galaxies in this sample seem
consistent with steep inner slopes, none of the galaxies {\it
require\/} halos with $\alpha=1$, as most galaxies are equally well or
better fit by halos with shallower density profiles or even constant
density cores.

Many of the systematic effects discussed in this paper can be avoided
if high-resolution, two-dimensional velocity fields are used. Slit
offsets will not play a role, the kinematical centers can be
determined directly from the data, and non-circular motions can be
mapped and modeled. A detailed study of barred dwarf and LSB galaxies
may be particularly interesting. 

\acknowledgements

We thank Vera  Rubin and Roelof Bottema for  providing useful comments
on earlier versions of this paper.   This research has made use of the
NASA/IPAC Extragalactic  Database (NED) which  is operated by  the Jet
Propulsion  Laboratory,  California  Institute  of  Technology,  under
contract with  the National Aeronautics and  Space Administration.  RS
thanks IPAC for its hospitality during his visits which were funded in
part  by  a  grant  to  BFM  as  part  of  the  NASA  Long-Term  Space
Astrophysics Program.

\end{document}